# Benchmarking Commercial Speech Recognition and Multimodal Large Language Models on Dysarthric Speech: Severity-Stratified Baselines and Architecture-Specific Prompting Effects

Ali Alsayegh[1]*, Tariq Masood[1]

A. Alsayegh, T. Masood

[1]Department of Design, Manufacturing and Engineering Management, University of Strathclyde, Glasgow, UK

Author Email Addresses:

1. *a.alsayegh@strath.ac.uk
2. tariq.masood@strath.ac.uk

Accepted Manuscript

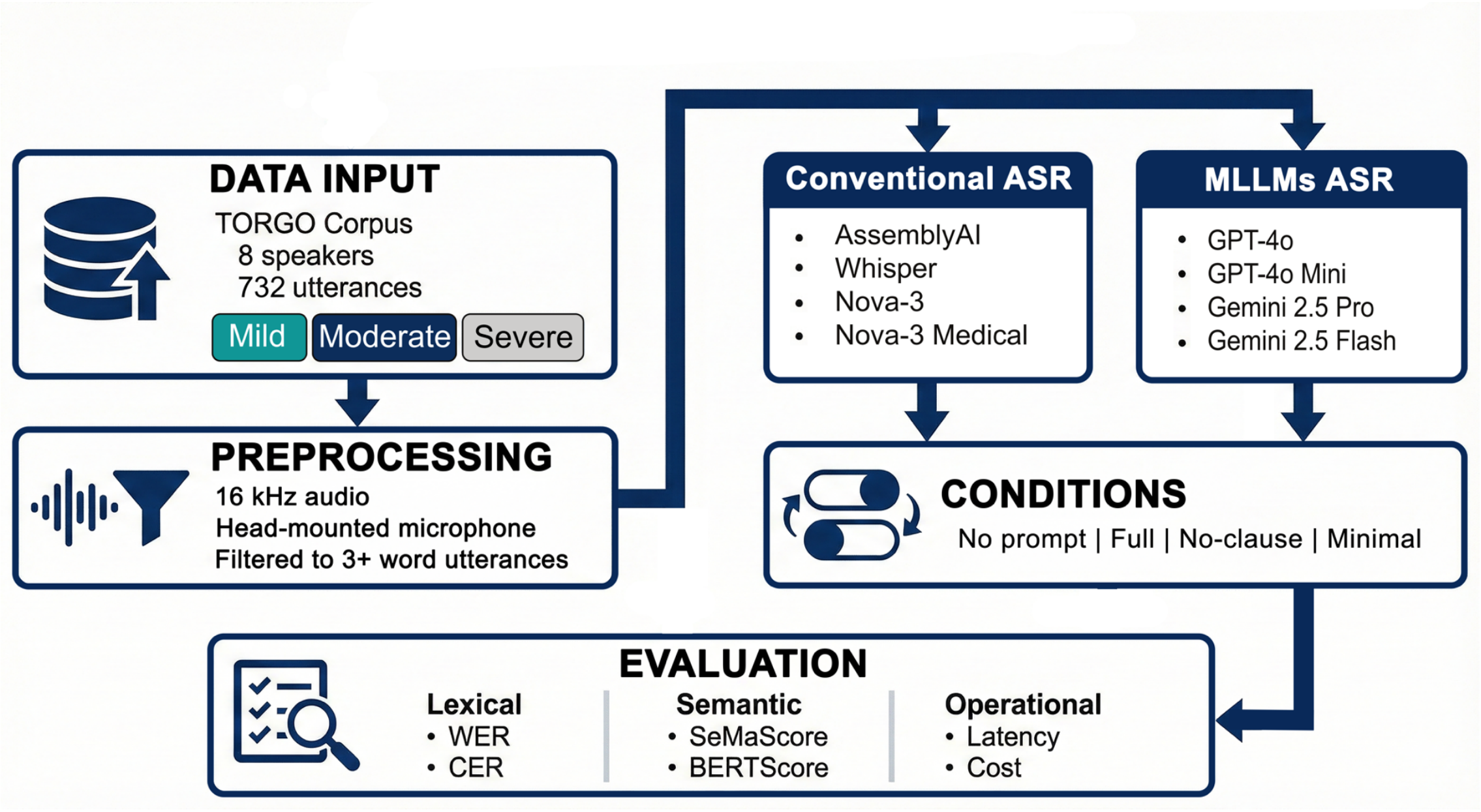

**Highlight:**

- Conducts the first systematic zero-shot evaluation of commercial ASR and multimodal large language models on dysarthric speech across severity levels
- Establishes severity-stratified performance baselines for eight commercial speech recognition services using the TORGO dysarthric speech corpus
- Reveals architecture-specific prompting effects where GPT-4o improves by 7.36 percentage points whilst Gemini variants exhibit degradation
- Demonstrates that semantic preservation metrics indicate partial recovery of communicative intent despite elevated lexical error rates
- Provides practical deployment guidance for assistive technology practitioners selecting speech recognition systems for motor speech disorders

**Abstract**

Voice-based human-machine interaction has become a primary means of accessing intelligent systems, yet individuals with dysarthria are systematically excluded by persistent gaps in recognition accuracy. Whilst automatic speech recognition (ASR) achieves word error rates (WER) below 5% on typical speech, performance degrades sharply for dysarthric speakers; and although multimodal large language models (MLLMs) might compensate for acoustic degradation through contextual reasoning, their zero-shot behaviour on such speech remains uncharacterised. To address this, we evaluate eight commercial speech-to-text services on the TORGO dysarthric speech corpus, comprising four conventional ASR systems (AssemblyAI, Whisper large-v3, Deepgram Nova-3, Nova-3 Medical) and four MLLM-based systems (GPT-4o, GPT-4o Mini, Gemini 2.5 Pro, Gemini 2.5 Flash), across lexical accuracy, semantic preservation, and cost-latency trade-offs. Recognition degraded consistently with severity: mild dysarthria reached low single-digit WER, around 1 to 2% for the leading systems, approaching typical-speech benchmarks, whereas severe dysarthria exceeded 51% WER for every system, with the MLLMs offering no advantage over conventional ASR at this tier under default settings. A four-condition prompt ablation revealed strongly architecture-specific effects, whereby, for the OpenAI multimodal models, any verbatim-transcription prompt reduced severe-tier WER mainly by suppressing non-target-language drift, lowering GPT-4o from 60.1% to 52.9% and GPT-4o Mini from 66.0% to roughly 55%, with the three prompt variants performing similarly; the Gemini models, which did not exhibit this failure mode to the same extent, showed no consistent benefit and sometimes degraded. The semantic metrics correlated strongly with WER and were largely redundant with it in aggregate, but isolated a high-error tail in which communicative intent was partly preserved despite poor lexical accuracy. We offer these severity-stratified, per-speaker baselines as a reusable reference for evidence-based technology selection in assistive voice interface deployment, rather than as a claim of system superiority.

# 1 Introduction

The principle of universal access posits that information society technologies should be accessible and usable by all citizens, including individuals with disabilities, without requiring specialised adaptation or a posteriori modification [1]. Voice-based interfaces have proliferated as a primary interaction modality, embedded within smart home devices, mobile assistants, automotive systems, and augmentative and alternative communication (AAC) platforms. For many users with motor impairments, voice control represents not merely a convenience but the most viable, or sole, means of interacting with digital technologies [2]. Yet the recognition accuracy that enables seamless voice interaction for typical speakers systematically fails for individuals with dysarthria, a motor speech disorder arising from neurological conditions including cerebral palsy, Parkinson's disease, amyotrophic lateral sclerosis, and stroke sequelae [3]. This recognition disparity creates a paradox wherein those who stand to benefit most from voice-mediated accessibility are precisely those for whom such interfaces remain functionally inaccessible.

The magnitude of this accessibility barrier is substantial. Dysarthria affects an estimated 170 per 100,000 individuals in the general population, with prevalence increasing to 70 to 100% among those with specific neurological conditions [3]. Contemporary automatic speech recognition (ASR) systems routinely achieve word error rates (WER) below 5% on typical speech [4], yet performance degrades precipitously for dysarthric speakers, frequently rendering voice interfaces non-functional for communicating user intent [5]. From the perspective of design for all, this performance disparity represents a significant challenge to inclusive design principles: commercial voice technologies are implicitly designed for a narrow band of typical speech characteristics, thereby excluding users whose speech patterns deviate from this assumed norm [6]. The consequences extend beyond individual inconvenience to systematic digital exclusion, reduced participation in the information society, and diminished autonomy for individuals who may already face communication barriers in their daily lives.

Critically, the deployment context for accessible speech recognition differs fundamentally from research settings. In practice, commercial speech-to-text services are deployed for clinical and assistive applications in zero-shot configurations: collecting speaker-specific adaptation data is frequently impractical, burdensome for users, or technically infeasible, and systems must function out of the box with no prior exposure to the individual's speech patterns [7]. This constraint reflects the reality faced by assistive technology practitioners and AAC developers who must select and configure recognition engines for diverse end-users. Moreover, evaluation of accessibility-relevant performance must extend beyond conventional lexical error metrics. For assistive applications, particularly AAC message generation, voice-controlled environmental systems, and telecare interfaces, semantic preservation may matter as much as verbatim accuracy for communicative function [8]. Deployment viability is further constrained by operational factors such as cost and latency [9].

Recent advances in multimodal large language models (MLLMs) have introduced architectures that integrate acoustic processing with powerful language modelling capabilities [10]. These systems may offer advantages for dysarthric speech by drawing on contextual inference to compensate for degraded acoustic evidence. However, whether such contextual reasoning translates to improved recognition of acoustically degraded speech remains an empirical question, and their effectiveness for dysarthric speech in particular is largely unexplored. The role of prompting, a technique widely used in text-based applications to shape model behaviour at inference, has not been systematically examined for speech transcription tasks involving atypical input.

To address these gaps, we present a comprehensive zero-shot evaluation of commercial speech recognition services on the TORGO dysarthric speech corpus [11]. Our study makes three primary contributions to the universal access literature. *First*, we establish severity-stratified and per-speaker baselines for eight commercial services on dysarthric speech, reporting lexical accuracy, semantic preservation, and operational cost and latency on a common footing. We position these as a reusable reference point rather than a claim of priority: they quantify how far current commercial systems remain from usable performance at each severity tier and expose the within-severity, between-speaker variability that aggregate figures conceal. *Second*, and most distinctively, we examine the effect of prompting through a four-condition ablation. We show that severe-speech improvements are architecture-specific: any verbatim-transcription prompt reduces severe-tier WER for the OpenAI multimodal models, mainly by suppressing non-target-language drift, whereas the effect is absent or inconsistent for the Gemini models, which do not exhibit the same degree of non-target-language drift. We also show that prompting can degrade performance on milder speech, most clearly for Gemini 2.5 Pro. This turns a single-prompt observation into a mechanism with a concrete deployment rule: prompt selectively, at severe dysarthria, and only for models that drift. *Third*, we examine what the semantic metrics add over WER. We quantify their relationship, finding a strong negative correlation with WER and near-redundancy between the two semantic measures, and we identify the high-error, intent-preserved tail in which lexical accuracy is poor yet communicative meaning is partly retained, illustrated with representative transcriptions. This clarifies when semantic evaluation is informative for assistive use and when it largely restates WER, tempering rather than overstating the case for meaning-based metrics. Throughout, we use the term zero-shot in the sense established in the speech and speech-LLM literature, to denote inference without task-specific fine-tuning or speaker adaptation, rather than the zero-shot-learning sense of recognising previously unseen classes [12]. No speaker-specific data, fine-tuning, or adaptation of any kind is used in this study.

## 2 Theoretical background and related work

### 2.1 Universal access and inclusive voice interface design

Universal access, as conceptualised within human-computer interaction research, encompasses the systematic effort to develop information technologies that are accessible, usable, and acceptable to the widest possible range of users across diverse abilities, contexts, and technological platforms [1, 13]. This paradigm extends beyond retrofitting accessibility features onto existing systems, characterised as reactive or accommodating approaches, to advocate for proactive design for all methodologies wherein diversity is treated as a fundamental design parameter from inception [14]. Voice interfaces present particular challenges for universal access: whilst they eliminate barriers associated with visual or manual interaction for some user groups, they simultaneously erect new barriers for users whose speech falls outside typical patterns.

The accessibility of voice interfaces thus exemplifies what Stephanidis and Savidis [13] term the interaction design challenge of universal access: developing systems that function effectively across the full spectrum of user diversity without requiring users to adapt to system limitations. For dysarthric speakers, current voice technologies fail this challenge: they demand either that users modify their speech to match system expectations (an impossibility for many dysarthric individuals) or that they abandon voice interaction entirely in favour of alternative modalities. This forced modality switching contradicts user-centred design principles and may impose cognitive and physical burdens that compound existing communication challenges [15]. The present study therefore addresses a critical gap in universal access research: characterising the current state of commercial speech recognition accessibility for dysarthric users and identifying pathways toward more inclusive voice interface design.

### 2.2 Dysarthria and the ASR accessibility gap

Dysarthria encompasses a heterogeneous group of motor speech disorders resulting from disruption to the neuromuscular control of speech production [3]. Acoustic manifestations include imprecise consonant articulation, vowel distortion, abnormal prosody (stress, intonation, and rhythm), hypernasality, and variable speech rate, characteristics that fundamentally violate the acoustic-phonetic assumptions underlying conventional ASR architectures [16]. Studies consistently report substantial recognition performance gaps: dysarthric speech recognition frequently exceeds 30% WER for moderate impairment and 60% WER for severe cases [5, 17, 18]. The heterogeneous aetiologies of dysarthria, spanning spastic, flaccid, ataxic, hypokinetic, hyperkinetic, and mixed presentations, further complicate recognition, as acoustic distortions differ qualitatively across subtypes [19].

Prior research has explored various approaches to bridging this accessibility gap. Speaker-adaptive training using techniques such as Maximum Likelihood Linear Regression (MLLR) has demonstrated relative WER reductions of approximately 37% [20], whilst speaker-dependent systems can achieve further improvements when sufficient adaptation data are available [21]. Self-supervised learning approaches including wav2vec 2.0 and HuBERT, pre-trained on large typical speech corpora, have shown promise for improving performance with limited dysarthric training data [22]. However, these approaches share a common limitation from a universal access perspective: they require speaker-specific data collection and model adaptation [23]. The present study addresses this limitation by evaluating systems under realistic zero-shot constraints.

## 2.3 Multimodal language models and context-aware transcription

The emergence of multimodal large language models represents an architectural evolution from purely acoustic-phonetic decoding toward context-aware transcription [10]. By integrating speech processing with powerful language modelling and extensive world knowledge, MLLMs may leverage semantic plausibility to disambiguate acoustically degraded input, potentially compensating for the phonetic confusion that characterises dysarthric speech. However, it remains unclear whether these systems' contextual reasoning capabilities actually improve recognition of acoustically degraded speech, or whether they primarily benefit from linguistic context in typical speech scenarios. Early evidence suggests such models demonstrate enhanced robustness to accented, noisy, and disfluent speech compared to conventional ASR [24, 25], though systematic evaluation on dysarthric speech remains limited.

Beyond speech processing, large language models have demonstrated capabilities in knowledge-intensive retrieval tasks, including extracting structured information from unstructured organisational data [26]. This capacity to bridge degraded or ambiguous input and coherent output is directly relevant to dysarthric speech recognition, where acoustic distortions must be resolved into accurate linguistic representations.

Prompting, the provision of natural language instructions to shape model behaviour at inference, has proven effective for many text-based tasks but remains underexplored for speech transcription. In this study, prompting is a textual instruction that constrains the multimodal model's transcription behaviour rather than a fused perceptual modality; it is therefore distinct from audio-visual fusion, in which complementary perceptual streams are combined within a single model. For dysarthric speech specifically, prompts might instruct models to expect atypical acoustic characteristics, prioritise verbatim transcription over grammatical correction, or tolerate phonetic ambiguity. Whether such instructions can improve recognition accuracy, maintain semantic fidelity, or alter the accuracy-latency trade-off remains an empirical question with direct implications for assistive technology deployment.

## 2.4 Evaluation frameworks for accessible speech technology

Traditional evaluation of ASR performance relies primarily on lexical error metrics, specifically WER and character error rate (CER) [27], that quantify surface-level discrepancies between hypothesis and reference transcriptions. WER is retained here as the principal lexical metric for comparability with prior dysarthric-ASR benchmarks, whilst recognising that character error rate has been argued to align more closely with human judgement than WER even for English [28]; the semantic measures introduced below complement, rather than replace, these word-level metrics. Whilst these metrics capture recognition accuracy in a technical sense, they inadequately represent communicative adequacy for assistive applications [8, 29]. A transcription that substitutes *medication* for *medicine* incurs the same WER penalty as one that substitutes *medication* for *celebration*, despite the former preserving semantic intent whilst the latter corrupts it entirely. For AAC applications, voice-controlled assistive devices, and telecare systems, conveying user intent matters more than verbatim accuracy.

Embedding-based semantic similarity measures including SeMaScore [8] and BERTScore [30] address this limitation by computing similarity in continuous representation space, capturing meaning preservation even when surface forms differ. We selected these two measures deliberately and as a complementary pair rather than as an exhaustive survey of semantic metrics. SeMaScore was designed specifically for ASR evaluation and validated against expert human judgements on atypical and noisy speech, combining a segment-level error rate with embedding similarity so that it remains sensitive to recognition errors that purely embedding-based scores can overlook [8]. BERTScore, by contrast, is the established general-purpose semantic-similarity baseline and has a direct precedent in disordered-speech ASR evaluation, where it has been shown to track human intelligibility judgements more closely than WER alone [31]; we therefore foreground it as our primary semantic metric for comparability with that prior work. The use of ASR-derived measures to index intelligibility in dysarthric speech has further been validated against gold-standard clinical ratings across multiple neurological conditions and languages [32], supporting the broader approach adopted here. Because the two measures map hypotheses to references differently, SeMaScore through Levenshtein-based segment alignment and BERTScore through token-level cosine similarity, agreement between them provides a more robust indication of preserved meaning than either alone. These metrics were nonetheless developed and validated primarily on typical speech transcription tasks; their behaviour on dysarthric speech, where both reference utterances and recognition hypotheses may contain atypical lexical patterns, has not been extensively validated.

## 3 Methodology

This study employed a comparative evaluation design assessing eight commercial speech-to-text services on dysarthric speech under zero-shot conditions. Fig. 1 summarises the experimental pipeline, implemented using the TranscribeSight benchmarking toolkit [33, 34], which comprised five stages: corpus preprocessing and speaker stratification, parallel API evaluation across conventional ASR and MLLM-based systems, prompting conditions for MLLMs, multi-dimensional metric computation, and severity-stratified statistical analysis.

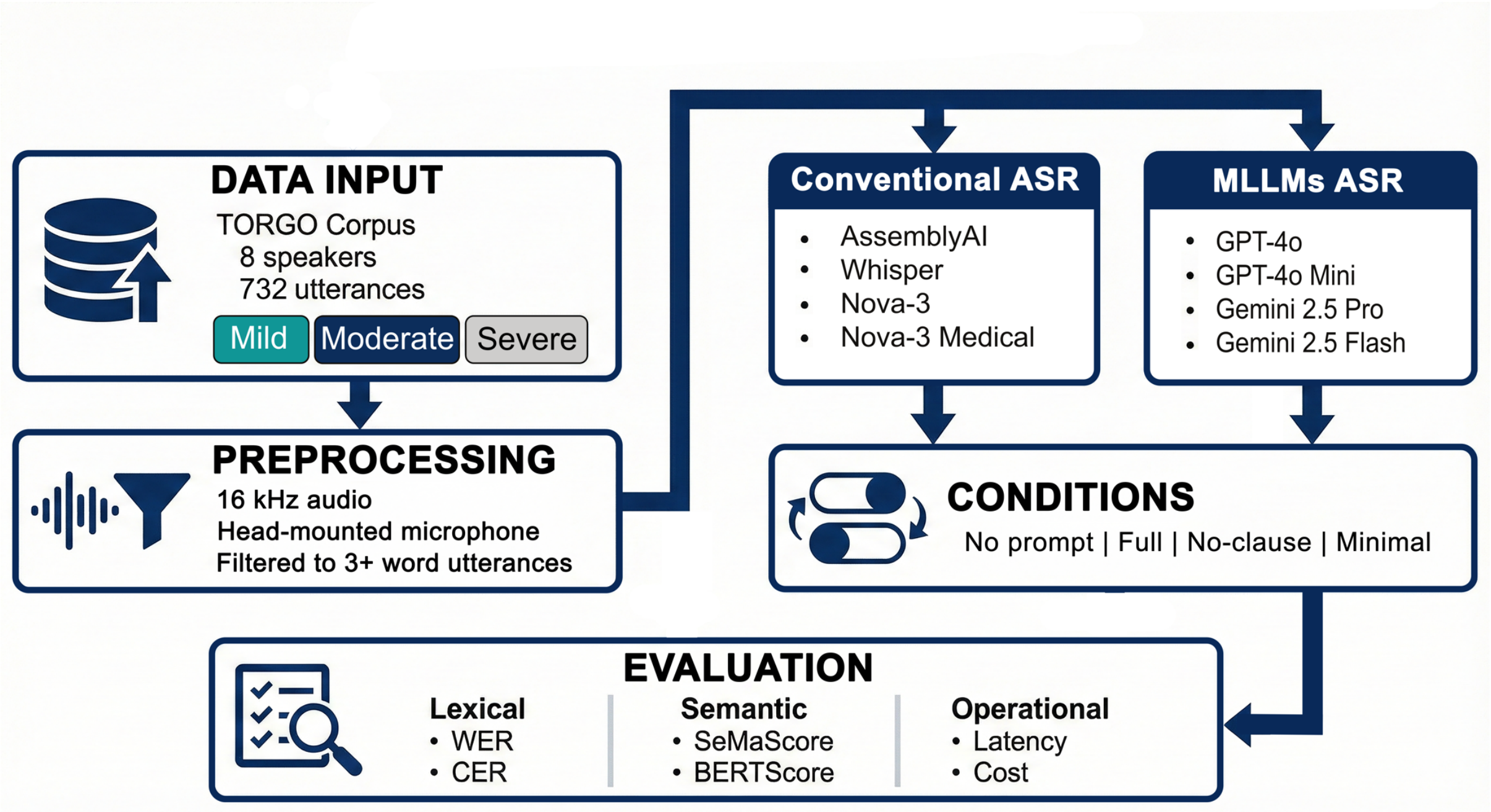

**Fig. 1** Experimental pipeline for comparative evaluation of speech-to-text services on dysarthric speech, implemented using the TranscribeSight benchmarking toolkit [33, 34]. The TORGO corpus (8 speakers, 732 utterances) was pre-processed and stratified by severity, then evaluated across eight commercial services: four conventional ASR systems (AssemblyAI, Whisper, Deepgram Nova-3, Nova-3 Medical) and four MLLM-based systems (GPT-4o, GPT-4o Mini, Gemini 2.5 Pro, Gemini 2.5 Flash). MLLM systems were evaluated under four prompt conditions: no prompt, full prompt, no-clause prompt, and minimal prompt. Multi-dimensional metrics (lexical, semantic, operational) were computed per utterance and analysed with severity stratification.

## 3.1 Corpus and participant characteristics

The study employed the TORGO corpus [11], a publicly available database of dysarthric and control speech designed for ASR research. TORGO was selected over alternative dysarthric speech corpora (e.g., UASpeech [35]) because its combination of connected speech utterances and head-mounted microphone recordings closely approximates realistic assistive technology deployment conditions; additionally, TORGO's established use in prior dysarthric ASR research facilitates comparison with existing literature. We prioritised evaluation depth across multiple commercial systems over breadth across corpora; the evaluation pipeline is designed for extension to additional datasets in future work.

TORGO contains recordings from eight speakers with dysarthria secondary to cerebral palsy (n = 5) or amyotrophic lateral sclerosis (n = 3), alongside seven age-matched and gender-matched control speakers. Recordings span single words, sentences, and paragraph-level utterances captured using both head-mounted and array microphones. We used the head-mounted microphone condition throughout, as it provides the most consistent low-noise signal across speakers and thereby isolates the effect of dysarthria on recognition from variation in recording conditions; it also corresponds most closely to the close-talking capture assumed by many assistive applications.

For this study, we restricted evaluation to the dysarthric subset and stratified analyses by the corpus-provided severity classifications: mild (speakers F04, M03), moderate (speakers F03, M05), and severe (speakers F01, M01, M02, M04). These severity labels are corpus-provided rather than derived from standardised clinical assessments such as the Frenchay Dysarthria

Assessment [36]; consequently, they should be interpreted as relative categories within this dataset rather than clinically validated severity levels.

It is important to note that dysarthria secondary to cerebral palsy and ALS may manifest differently in terms of acoustic characteristics. Cerebral palsy typically produces spastic or mixed dysarthria with relatively stable speech patterns, whilst ALS-related dysarthria often involves progressive deterioration and may include flaccid components. The TORGO corpus does not enable systematic comparison across aetiologies due to sample size constraints, and our findings should be interpreted as characterising recognition performance across this mixed aetiology sample rather than generalising to specific dysarthria subtypes.

Control speaker recordings were excluded from analysis as the primary research question concerns the accessibility gap for dysarthric users rather than general ASR benchmarking. Typical-speech recognition is already well-characterised in existing benchmarks, and including controls would not inform assistive deployment decisions where the critical question is absolute performance on atypical speech rather than relative degradation from a typical baseline. Pipeline validation on typical speech is reported in our prior work [33], where we benchmarked eight ASR services on English LibriSpeech samples using default API configurations within the TranscribeSight framework; the present study focuses on the accessibility gap for dysarthric users.

A three-stage inclusion protocol was applied to ensure evaluation focused on phrase-level recognition relevant to assistive applications: (i) utterances with fewer than three words were excluded to focus on connected speech, since such single- and two-word items (for example, isolated responses such as yes, no, or stop) do not exercise the connected-speech recognition that is the focus of this study; (ii) items lacking reference transcriptions were removed; and (iii) non-lexical or instructional prompts were discarded.

Scoring-stage text normalisation of the reference and hypothesis transcripts is described in Section 3.4. Table 1 summarises the final dataset, comprising 732 utterances distributed across severity levels. This provided sufficient data for descriptive severity-stratified comparison whilst maintaining ecological validity for assistive deployment scenarios.

The excluded utterances were predominantly single words and short phrases from elicitation tasks. Their exclusion may limit the generalisability of the findings to single-word command-recognition scenarios, which are common in some AAC and smart-home applications. However, such command-and-control systems often employ constrained vocabularies and grammar-based recognition approaches, which may produce different recognition patterns from the open-vocabulary evaluation conducted here. Future research should therefore specifically address short-command recognition using appropriately designed evaluation protocols.

**Table 1** Distribution of included TORGO speakers and utterances by dysarthria severity.

| Severity | Speaker | Initial utterances | Final utterances |
|---|---|---|---|
| Mild | F04 | 431 | 100 |
| | M03 | 407 | 95 |
| Moderate | F03 | 546 | 138 |
| | M05 | 502 | 115 |
| Severe | F01 | 117 | 20 |
| | M01 | 371 | 89 |
| | M02 | 389 | 91 |
| | M04 | 391 | 84 |
| **Total** | | **3,154** | **732** |

### 3.2 Systems evaluated

Eight commercially available speech-to-text services were evaluated, representing the current landscape of technologies potentially deployable for assistive applications. *Conventional ASR systems* included AssemblyAI (default model), OpenAI Whisper large-v3 [37] via API, Deepgram Nova-3, and Deepgram Nova-3 Medical (a variant optimised for healthcare terminology). *MLLM-based services* included OpenAI GPT-4o (gpt-4o-audio), OpenAI GPT-4o Mini, Google Gemini 2.5 Pro, and Google Gemini 2.5 Flash. All systems were accessed via their public APIs during a single evaluation window (17 to 18 September 2025) to ensure temporal consistency in model versions.

Audio data were sourced exclusively from the TORGO wav_headMic recordings, which were captured using a head-mounted electret microphone at an original sampling rate of 22.1 kHz, subsequently downsampled to 16 kHz for distribution. Audio preprocessing was minimal to reflect realistic deployment conditions: recordings were converted to 16 kHz mono WAV format where required by API specifications, with no noise reduction, normalisation, or enhancement applied. Some TORGO recordings affected by electromagnetic interference from the articulograph system were excluded from the corpus by its creators; the remaining recordings represent typical quality achievable in controlled recording environments. Critically, all systems were assessed in a zero-shot configuration: no task-specific training, fine-tuning, or speaker adaptation was performed. Conventional ASR systems were invoked with default API parameters. This evaluation design reflects realistic deployment constraints for assistive technology applications where system configuration must generalise across diverse end-users without per-user customisation.

For reproducibility, we report the specific model identifiers and API configurations used: OpenAI Whisper large-v3 (whisper-1 endpoint), GPT-4o (gpt-4o-audio, September 2025 version), GPT-4o Mini (gpt-4o-mini-audio), AssemblyAI (default model via v2 API), Deepgram Nova-3 and Nova-3 Medical (nova-3 and nova-3-medical models), and Google Gemini 2.5 Pro and Flash (gemini-2.5-pro and gemini-2.5-flash). The settings most likely to affect results were held constant across all eight systems. Decoding strategy and sampling temperature were left at provider defaults and were not tuned per system, and all audio was supplied as 16 kHz mono with no noise reduction, normalisation, or enhancement. This means any future divergence from the results reported here is attributable to provider-side model or default changes rather than to configuration choices made in this study.

This evaluation should be understood as a time-stamped benchmark representing the state of commercial speech recognition practice as of 17-18 September 2025. Results may differ with subsequent model updates, as commercial APIs are subject to periodic revision without version guarantees; practitioners should verify current performance before deployment decisions. The evaluation pipeline was implemented using TranscribeSight [33, 34], an open-source benchmarking toolkit; code and configuration files are available in the linked repository to facilitate replication and longitudinal tracking of API evolution.

### 3.3 Prompting protocol for MLLM systems

MLLM-based services were evaluated under four conditions: a no-prompt baseline and three prompted variants within a single verbatim-transcription strategy. This design was used to examine both the overall effect of prompting and the contribution of individual prompt components. The no-prompt condition reflected default model behaviour, whereas the prompted conditions were designed to test whether textual instructions could reduce over-correction, paraphrasing, grammatical normalisation, or non-target-language drift when processing dysarthric speech. The exact text of each condition is given in Table 2.

Each prompt was supplied once per request. No few-shot examples or additional system instructions were provided, and where temperature or decoding parameters were configurable, provider defaults were retained. Single API calls were made per utterance. The four evaluated conditions were: no prompt, full prompt, no-clause prompt, and minimal prompt. The no-clause condition was identical to the full prompt except that the anti-inference sentence was removed, allowing the effect of that specific instruction to be isolated. The minimal condition reduced the instruction to a short verbatim-transcription directive.

We did not test broader prompting approaches such as few-shot prompting, prompt tuning, domain-specific vocabulary hints, or multi-turn correction, which have been explored in broader LLM and speech-language-model research [38]. More reasoning-oriented strategies, such as chain-of-thought prompting or instruction decomposition, were also not examined because verbatim transcription is primarily a perception-bound task rather than a multi-step reasoning task.

This decision reflects the perception-bound nature of verbatim transcription, where the primary task is to recover the spoken signal rather than to perform multi-step reasoning. These alternative strategies remain important directions for future work, but the present analysis focuses specifically on whether prompt wording within a verbatim-transcription strategy changes recognition behaviour.

**Table 2** Prompt conditions evaluated for the MLLM-based systems.

| Condition | Prompt supplied to the model |
|---|---|
| No prompt | None. The audio is submitted with no textual instruction, reflecting default system behaviour. This condition is the baseline reported for all eight systems. |
| Full | You are tasked with transcribing English dysarthric speech. The speech input may be slurred, slow, or imprecise. Produce a verbatim transcript in English, preserving the spoken output without correction, normalisation, or substitution. In cases where words are unclear, render them as closely as possible to the audible signal rather than inferring or guessing. Output requirements: Provide only the plain transcript text. Do not include labels, headings, explanations, or formatting of any kind. |
| No-clause | Identical to the Full prompt, with a single sentence removed: the anti-inference instruction "In cases where words are unclear, render them as closely as possible to the audible signal rather than inferring or guessing." This deletion is the only difference between the Full and No-clause conditions. |
| Minimal | Transcribe the English speech verbatim. Provide only the plain transcript text, with no correction, normalisation, explanation, or formatting. |

## 3.4 Evaluation metrics

*Lexical accuracy* was assessed using WER and CER, computed as the sum of substitution, deletion, and insertion errors divided by reference length. All transcripts were scored after applying the Whisper English text normaliser [37] to both the reference and the hypothesis. This published normaliser lowercases the text, removes punctuation, expands contractions, for example mapping “he’s” to “he is”, and converts spelled-out numbers to digits, for example mapping “ninety-three” to “93”. Applying the same normalisation symmetrically to the reference and the hypothesis means that differences in case, punctuation, contraction and number formatting do not count as recognition errors, whereas genuine lexical substitutions are still penalised. This procedure prevents superficial formatting differences from inflating WER while retaining penalties for genuine lexical errors, as illustrated in Table 6. As a robustness check, re-scoring under an alternative normalisation that spells numbers out rather than digitising them changed WER by at most 0.7 percentage points at any severity level and did not change the ranking of systems, including AssemblyAI remaining the lowest-WER system at severe dysarthria.

*Semantic fidelity* was evaluated using SeMaScore [8] and BERTScore [30], embedding-based measures that compute similarity between hypothesis and reference in continuous semantic space. SeMaScore was computed using the all-MiniLM-L6-v2 sentence transformer model, whilst BERTScore used roberta-large embeddings; both configurations represent commonly used defaults for these metrics. Following the multi-dimensional evaluation frameworks discussed in Section 2.4, we include these semantic alternatives alongside WER and CER to capture meaning preservation even when surface forms differ. Operational efficiency was characterised by cost (US dollars per input minute) and latency (API round-trip time from request initiation to response receipt). We acknowledge that latency measurements may be influenced by network conditions; all API calls were made from the same network location during the evaluation window to minimise this variability.

## 3.5 Aggregation and statistical reporting

All evaluations were conducted using TranscribeSight [33, 34], an open-source benchmarking toolkit implementing standardised scoring protocols. Metrics were computed per utterance; severity-stratified results are reported as mean plus or minus standard deviation.

For the group-level results reported in Table 3, WER is computed as the mean of per-utterance WER across all utterances within a severity stratum, with each utterance weighted equally and per-utterance WER capped at 0 to 100% before averaging; we refer to this as the utterance-weighted value. Utterances for which a system returned an empty transcription were scored as 100% WER, since an empty hypothesis deletes every reference word; this affected 14 utterances, all from the two Deepgram systems at severe dysarthria.

BERTScore is computed with roberta-large using baseline rescaling, which maps a perfect match to 100 and places scores on a nominal 0 to 100 scale. Unlike WER and CER, the embedding-based semantic metrics are reported on their native scales without clipping. For BERTScore, baseline rescaling means that a near-total mismatch can yield a value below 0, and we retain these values rather than flooring them, so that the reported means and the per-utterance examples reflect the true distribution of semantic similarity, including the strongly negative scores produced by non-target-language and degenerate outputs at severe dysarthria.

For the per-speaker results reported in Table 4, WER is computed in the same manner over each speaker's utterances, and where a speaker-weighted value is reported, each speaker contributes equally regardless of utterance count. Per-speaker values in Table 4 are reported as mean ± SD.

We use the terms utterance-weighted and speaker-weighted rather than pooled-word WER because the analysis focuses on per-utterance accessibility performance and prevents extreme generated outputs from dominating aggregate results. The two schemes agree closely except where a speaker with few utterances has atypical WER; for GPT-4o at severe dysarthria, the utterance-weighted mean is 60.1% and the speaker-weighted mean is 60.7%, with the small difference reflecting the unequal utterance counts across speakers, in particular speaker F01, who contributes only 20 utterances.

To assess the effect of prompting on the multimodal systems, we adopt a four-condition design comprising the no-prompt, full, no-clause and minimal conditions defined in Section 3.3. In the analyses that follow, references to the OpenAI multimodal models concern GPT-4o and GPT-4o Mini; Whisper, although also developed by OpenAI, is a conventional ASR system and was evaluated only in the non-prompted condition. We report condition-level mean WER and BERTScore by model and severity. Because the severe subset comprises only four speakers and the pooled per-utterance observations are not independent across systems, we treat these comparisons as descriptive and do not apply null-hypothesis significance testing. The non-prompted severe set comprises 284 utterances; of these, 279 are used where a directly paired per-utterance comparison between conditions is required.

# 4 Results

## 4.1 Severity-dependent performance degradation

Recognition performance showed a clear severity-dependent decline across all evaluated systems and metrics. As summarised in Table 3, lexical and semantic performance deteriorated progressively from mild through moderate to severe dysarthria under default, non-prompted configurations. For reference, the sub-5% typical-speech benchmarks noted in Section 1 (on corpora such as LibriSpeech [39] and Common Voice [40]) provide the implicit baseline against which these dysarthric-speech results should be interpreted. As noted in Section 3.1, the severity categories are dataset-relative labels provided by the TORGO corpus rather than clinically validated classifications; however, they represent consistent intelligibility tiers within the dataset and provide a meaningful basis for stratified accessibility evaluation.

This graded degradation confirms the expected relationship between speech impairment severity and recognition difficulty and, critically from a universal access perspective, demonstrates that individuals with greater communication needs face proportionally larger accessibility barriers. Fig. 2 visualises this lexical degradation across WER and CER.

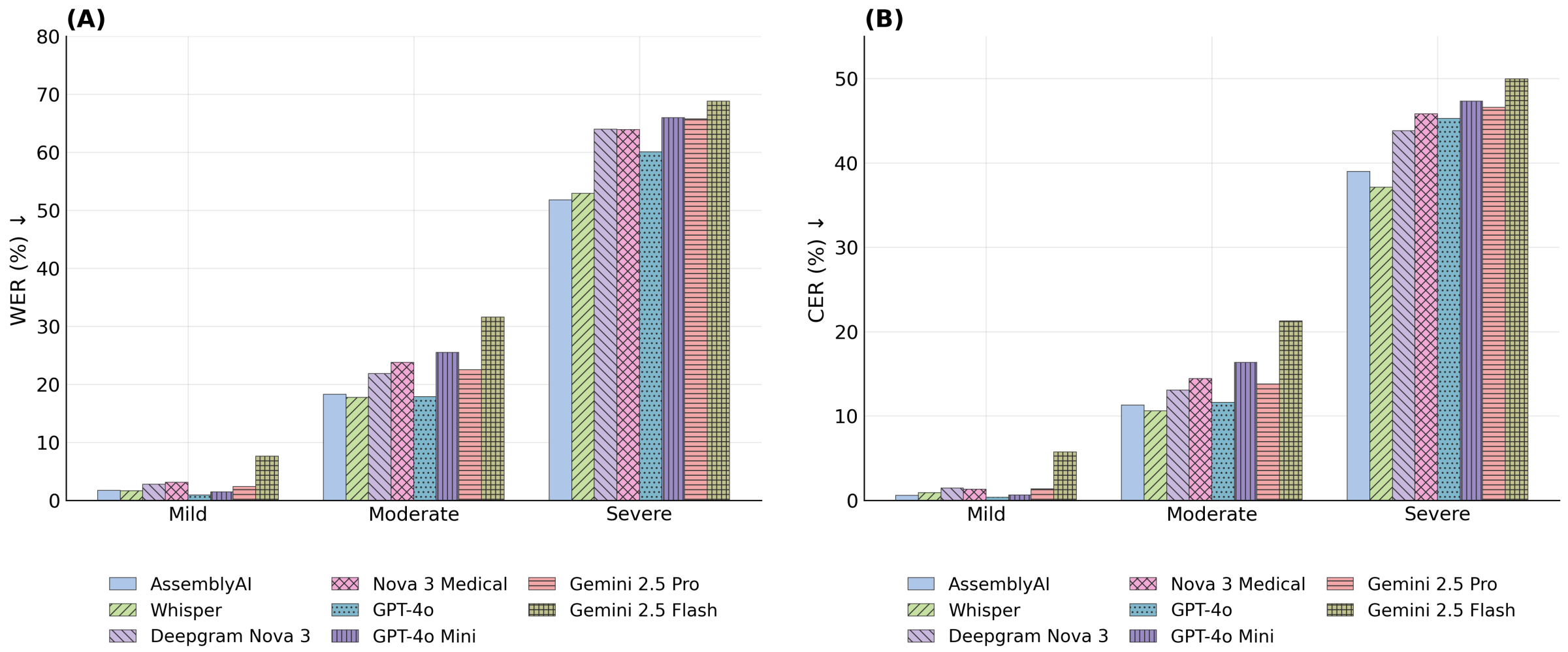

**Fig. 2** Lexical error across dysarthria severities under default configurations. Bars show mean WER (A) and CER (B) for conventional ASR systems (Deepgram Nova-3, Nova-3 Medical, AssemblyAI, Whisper) and MLLM systems (GPT-4o, GPT-4o Mini, Gemini 2.5 Pro, Gemini 2.5 Flash). Lower values indicate better performance.

For *mild dysarthria*, system performance clustered within operationally acceptable ranges. GPT-4o achieved the lowest WER (0.9%), followed by GPT-4o Mini (1.4%). Among the conventional ASR systems, Whisper (1.6%) and AssemblyAI (1.7%) were closely matched. Semantic preservation was high across systems (GPT-4o: 99.3% SeMaScore, 99.1% BERTScore), indicating that even when lexical errors occur, semantic intent is largely preserved. These results suggest that current commercial systems may provide adequate voice interface accessibility for users with mild dysarthria. It should be noted that these relatively low WER values reflect connected speech performance on utterances of three or more words; single-word command recognition, which was excluded from evaluation, may exhibit different accuracy characteristics.

For *moderate dysarthria*, performance declined substantially across all systems. Whisper achieved the lowest WER among conventional ASR (17.7%), whilst GPT-4o led among the MLLMs (17.9%). Semantic metrics remained above 80% SeMaScore for leading systems, suggesting partial preservation of communicative function despite elevated lexical error rates. Notably, Deepgram Nova-3 Medical, ostensibly optimised for clinical contexts, underperformed the standard Nova-3 variant (23.8% vs 21.8% WER). This unexpected finding may reflect the specific nature of medical vocabulary optimisation, which targets terminology rather than acoustic robustness to atypical speech; however, further investigation would be needed to draw definitive conclusions about domain-specific optimisation strategies.

For *severe dysarthria*, all systems exhibited substantial degradation, with WER exceeding 51% across the board. AssemblyAI achieved the lowest WER (51.8%), followed by Whisper (52.9%). MLLM-based systems, despite their contextual reasoning capabilities, did not outperform conventional ASR at this severity level under default configurations. Semantic preservation also dropped substantially, with SeMaScore ranging from 39% to 58%. This indicates a marked reduction in meaning preservation and suggests that many severe-dysarthria outputs may no longer support reliable communicative intent. Fig. 3 shows the corresponding decline in BERTScore and SeMaScore across severity levels.

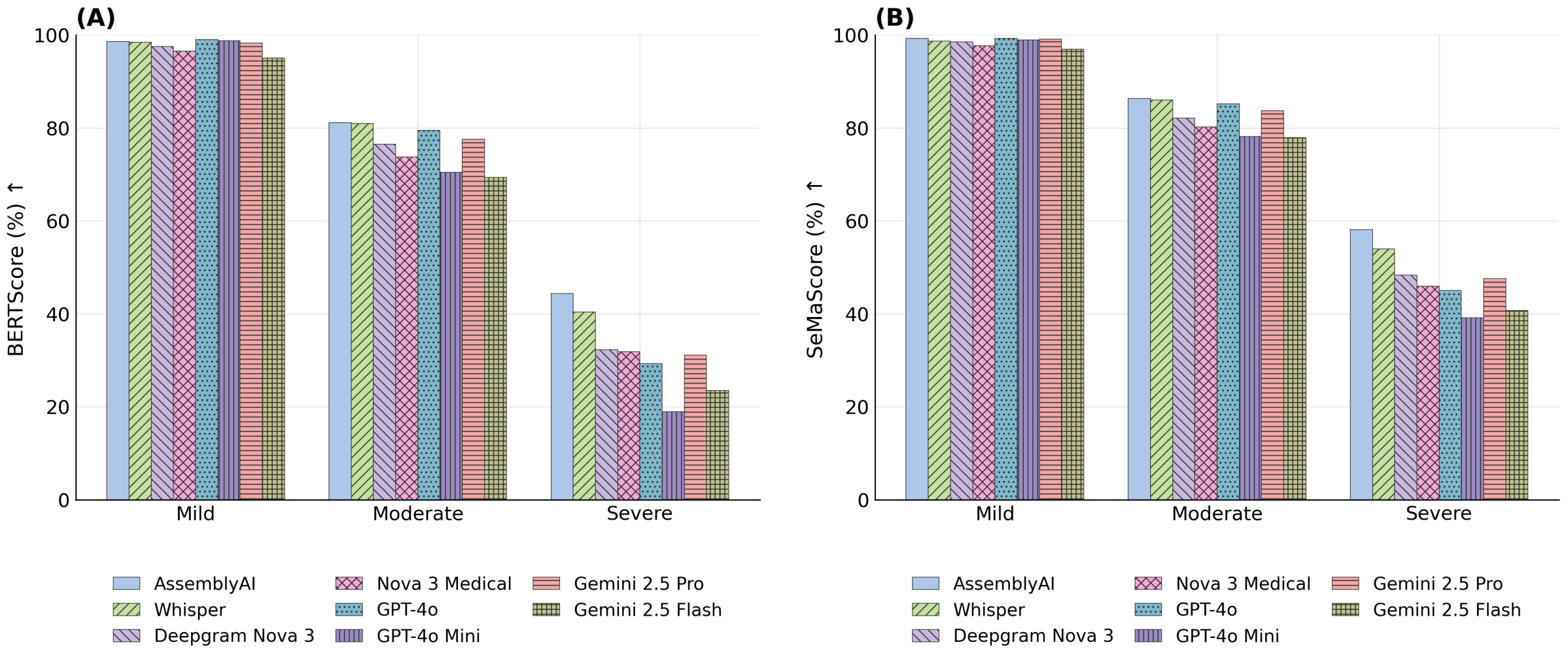

**Fig. 3** Semantic preservation across dysarthria severities under default configurations. Bars show mean BERTScore (A) and SeMaScore (B) for all systems. Higher values indicate greater semantic preservation. Results degrade from mild to severe dysarthria.

**Table 3** Performance of ASR and MLLM systems on TORGO under default (non-prompted) settings, reported as mean ± SD. WER values are utterance-weighted, with per-utterance WER capped at 100% before averaging.

| Severity | System | WER (%) ↓ | CER (%) ↓ | SeMaScore ↑ | BERTScore ↑ |
|---|---|---|---|---|---|
| Mild | AssemblyAI | 1.70 ± 5.43 | 0.57 ± 1.97 | 99.23 ± 3.06 | 98.63 ± 5.31 |
| | Whisper | 1.64 ± 5.57 | 0.88 ± 3.53 | 98.73 ± 5.45 | 98.47 ± 5.96 |
| | Deepgram Nova 3 | 2.77 ± 7.81 | 1.44 ± 5.42 | 98.54 ± 6.14 | 97.53 ± 8.02 |
| | Nova 3 Medical | 3.08 ± 8.21 | 1.30 ± 3.87 | 97.74 ± 8.41 | 96.56 ± 10.46 |
| | GPT-4o | **0.87 ± 3.65** | **0.35 ± 1.63** | **99.28 ± 4.02** | **99.06 ± 5.08** |
| | GPT-4o Mini | 1.40 ± 4.86 | 0.62 ± 2.84 | 98.93 ± 5.53 | 98.74 ± 4.68 |
| | Gemini 2.5 Pro | 2.34 ± 7.29 | 1.33 ± 5.58 | 99.09 ± 3.48 | 98.24 ± 5.92 |
| | Gemini 2.5 Flash | 7.56 ± 22.10 | 5.74 ± 20.39 | 96.92 ± 10.81 | 95.07 ± 13.41 |
| Moderate | AssemblyAI | 18.23 ± 23.54 | 11.27 ± 17.88 | **86.35 ± 19.59** | **81.13 ± 24.39** |
| | Whisper | **17.72 ± 25.26** | **10.62 ± 17.57** | 86.01 ± 22.63 | 80.93 ± 28.68 |
| | Deepgram Nova 3 | 21.83 ± 25.44 | 13.04 ± 17.19 | 82.15 ± 22.44 | 76.50 ± 27.51 |
| | Nova 3 Medical | 23.75 ± 27.72 | 14.41 ± 19.30 | 80.21 ± 24.09 | 73.78 ± 29.80 |
| | GPT-4o | 17.85 ± 27.87 | 11.58 ± 19.92 | 85.20 ± 25.97 | 79.47 ± 34.80 |
| | GPT-4o Mini | 25.45 ± 32.83 | 16.32 ± 22.65 | 78.16 ± 29.98 | 70.43 ± 39.62 |
| | Gemini 2.5 Pro | 22.50 ± 28.81 | 13.80 ± 19.63 | 83.68 ± 24.43 | 77.61 ± 30.10 |
| | Gemini 2.5 Flash | 31.57 ± 33.46 | 21.25 ± 26.35 | 77.88 ± 26.31 | 69.34 ± 32.67 |
| Severe | AssemblyAI | **51.77 ± 32.20** | 38.96 ± 28.76 | **58.07 ± 28.68** | **44.35 ± 34.44** |
| | Whisper | 52.94 ± 35.78 | **37.11 ± 29.06** | 53.95 ± 32.67 | 40.42 ± 41.27 |
| | Deepgram Nova 3 | 64.00 ± 31.79 | 43.82 ± 25.89 | 48.31 ± 28.47 | 32.27 ± 32.75 |
| | Nova 3 Medical | 63.89 ± 29.97 | 45.79 ± 25.64 | 45.91 ± 27.54 | 31.89 ± 31.21 |
| | GPT-4o | 60.08 ± 36.07 | 45.28 ± 29.03 | 45.00 ± 35.22 | 29.29 ± 47.57 |
| | GPT-4o Mini | 66.00 ± 36.02 | 47.32 ± 27.25 | 39.12 ± 34.72 | 18.97 ± 48.12 |
| | Gemini 2.5 Pro | 65.74 ± 35.10 | 46.58 ± 28.88 | 47.57 ± 32.38 | 31.10 ± 38.24 |
| | Gemini 2.5 Flash | 68.83 ± 31.39 | 49.96 ± 27.17 | 40.71 ± 29.44 | 23.51 ± 34.31 |

## 4.2 Per-speaker analysis and speaker-weighted results

To examine robustness of findings across individual speakers, Table 4 presents per-speaker WER for all systems under non-prompted conditions. This analysis reveals substantial within-severity variability, particularly for severe dysarthria. Among severe speakers, M02 achieved the lowest mean WER across systems (49.3%), whilst M04 exhibited the highest (77.5%), a difference of over 28 percentage points despite both being classified as severe. This variability more plausibly reflects speaker-specific acoustic and articulatory differences, together with differences in error behaviour and the underlying distribution of per-utterance scores.

Two features of the TORGO corpus qualify the per-speaker comparison. First, the number of utterances available per speaker is unequal, ranging from 20 to 91 at severe dysarthria, so per-speaker means are estimated with differing precision and the speakers with fewer utterances yield less stable estimates. Second, TORGO does not present an identical prompt set to every speaker, so part of the between-speaker variation may reflect differences in the spoken content rather than severity alone. We therefore treat the per-speaker values as descriptive and place greater weight on the severity-level comparisons, which aggregate over speakers and prompts and are correspondingly less sensitive to any single speaker or prompt, although they too remain descriptive given that the prompt set is not fully balanced across speakers.

Comparison of speaker-weighted (per-speaker mean) and utterance-weighted (per-utterance) results revealed generally consistent patterns, with differences below approximately 1.6 percentage points across systems at severe dysarthria. For GPT-4o, the speaker-weighted analysis yielded 60.7% WER compared with 60.1% in the utterance-weighted analysis. These findings suggest that severity-level conclusions are robust to the choice of averaging scheme (utterance- versus speaker-weighting, defined in Section 3.5), though practitioners should recognise that individual speaker performance may deviate substantially from severity-level averages.

**Table 4** Per-speaker WER (%) under non-prompted conditions. Speakers are grouped by severity; n indicates utterance count per speaker. Cells report mean ± SD.

| Severity | Speaker | n | Asm | Whsp | DG | DGM | GPT | Mini | Pro | Flash |
|---|---|---|---|---|---|---|---|---|---|---|
| Mild | F04 | 100 | 1.7 ± 5.3 | 1.6 ± 5.0 | 3.5 ± 8.7 | 3.8 ± 9.2 | 0.9 ± 3.4 | 1.6 ± 5.2 | 2.8 ± 8.2 | 8.2 ± 22.9 |
| | M03 | 95 | 1.7 ± 5.6 | 1.6 ± 6.2 | 2.0 ± 6.7 | 2.3 ± 7.0 | 0.9 ± 3.9 | 1.2 ± 4.5 | 1.9 ± 6.3 | 6.9 ± 21.3 |
| Moderate | F03 | 138 | 10.5 ± 16.3 | 13.5 ± 22.7 | 13.1 ± 19.7 | 15.8 ± 22.7 | 12.4 ± 24.1 | 17.5 ± 28.8 | 16.0 ± 25.3 | 25.5 ± 32.9 |
| | M05 | 115 | 27.5 ± 27.3 | 22.8 ± 27.3 | 32.3 ± 27.6 | 33.3 ± 30.2 | 24.4 ± 30.6 | 35.0 ± 34.9 | 30.3 ± 30.9 | 38.9 ± 32.8 |
| Severe | F01 | 20 | 48.1 ± 25.9 | 53.8 ± 30.5 | 54.5 ± 25.2 | 64.9 ± 20.1 | 61.8 ± 35.7 | 72.5 ± 32.9 | 59.5 ± 37.3 | 60.2 ± 30.9 |
| | M01 | 89 | 56.1 ± 31.4 | 49.0 ± 34.8 | 62.7 ± 29.6 | 61.6 ± 28.5 | 58.0 ± 36.4 | 62.5 ± 35.2 | 63.7 ± 33.5 | 65.2 ± 30.2 |
| | M02 | 91 | 36.3 ± 33.6 | 41.0 ± 36.2 | 53.4 ± 35.6 | 54.7 ± 34.2 | 47.7 ± 37.7 | 53.0 ± 39.8 | 50.2 ± 37.7 | 57.7 ± 35.2 |
| | M04 | 84 | 64.8 ± 25.5 | 69.9 ± 31.3 | 79.1 ± 24.9 | 76.0 ± 24.4 | 75.4 ± 28.0 | 82.2 ± 25.9 | 86.2 ± 21.2 | 86.7 ± 18.4 |

*Note: Asm = AssemblyAI, Whsp = Whisper, DG = Deepgram Nova 3, DGM = Deepgram Nova 3 Medical, GPT = GPT-4o, Mini = GPT-4o Mini, Pro = Gemini 2.5 Pro, Flash = Gemini 2.5 Flash*

The large per-speaker standard deviations in Table 4 are best explained not by a few isolated outliers, but by a near-bimodal distribution of per-utterance WER. Fig. 4 shows that, because per-utterance WER is capped between 0 and 100%, scores accumulate at the two extremes. At severe dysarthria, 9.9% of utterances were recognised perfectly (WER = 0), whereas 27.8% reached the 100% ceiling, with a median WER of 67%. At mild dysarthria, the distribution was instead concentrated near zero, with 87.2% of utterances at WER = 0 and only 0.6% at the ceiling. The moderate tier fell between these two patterns, with 42.7% at zero and 5.4% at the ceiling, and a median WER of 12%. This bimodal structure, rather than a small number of extreme values, explains the wide per-speaker standard deviations. Fig. 5 shows that this near-bimodal shape is present for every system rather than being driven by a subset, with the proportion of severe utterances at the 100% ceiling ranging from about 15% for AssemblyAI to about 38% for GPT-4o Mini.

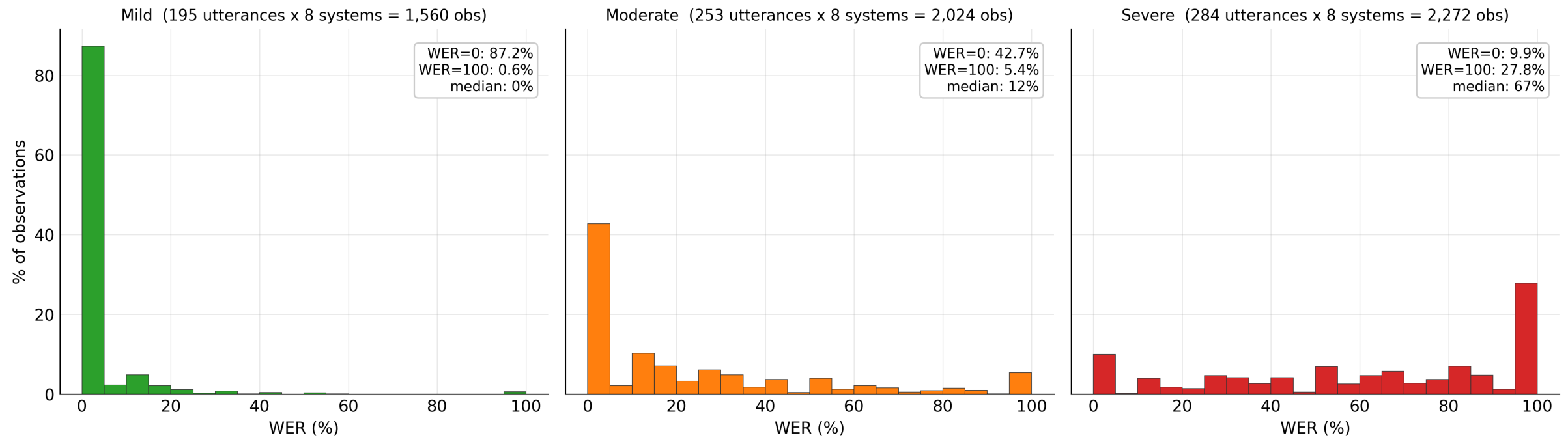

**Fig. 4** Per-utterance WER distribution by severity under the no-prompt baseline, pooled across the eight systems. Panels are labelled with the underlying utterance and observation counts, corresponding to the severity strata in Table 1. The distribution is concentrated near zero at mild dysarthria and becomes near-bimodal at severe dysarthria, with mass at both the 0% and 100% extremes.

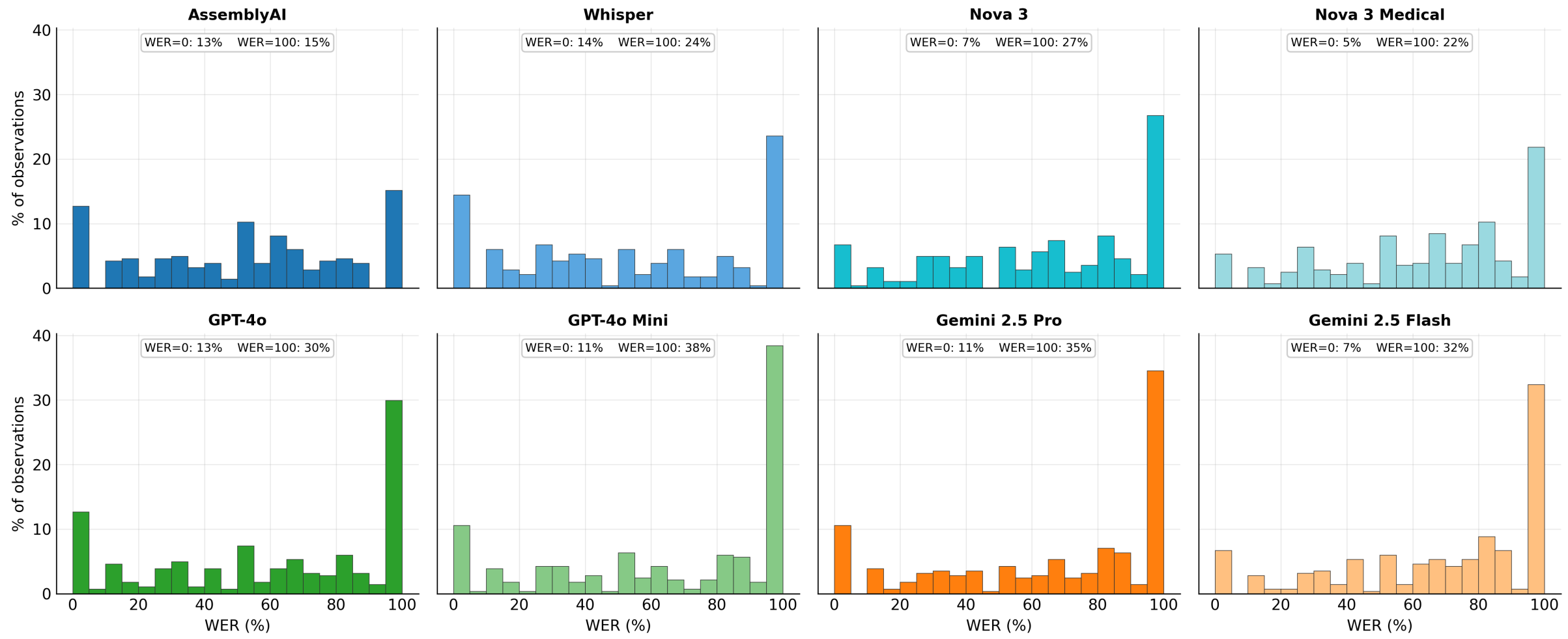

**Fig. 5** Per-utterance WER distribution at severe dysarthria, shown separately for each of the eight systems under the no-prompt baseline. The near-bimodal shape seen in the pooled distribution in Fig. 4 is present for every system, with mass at both the 0% floor and the 100% ceiling rather than being driven by a subset of systems. The proportion of utterances at the 100% ceiling ranges from about 15% for AssemblyAI to about 38% for GPT-4o Mini.

As shown in Table 5, utterance length was comparable across speakers, with mean utterance length ranging from approximately seven to nine words and individual utterances spanning three to fourteen words in every severity tier. The per-speaker WER differences reported above are therefore more likely to reflect speaker- and severity-specific recognition effects than systematic differences in utterance length.

**Table 5** Reference-transcript words per utterance by speaker.

| Severity | Speaker | n | Mean | Median | Min | Max |
|---|---|---|---|---|---|---|
| mild | F04 | 100 | 7.4 | 7.0 | 4 | 14 |
| mild | M03 | 95 | 7.5 | 7.0 | 4 | 14 |
| moderate | F03 | 138 | 7.3 | 7.0 | 3 | 14 |
| moderate | M05 | 115 | 7.3 | 7.0 | 4 | 14 |
| severe | F01 | 20 | 9.2 | 8.5 | 4 | 14 |
| severe | M01 | 89 | 7.3 | 7.0 | 3 | 14 |
| severe | M02 | 91 | 7.3 | 7.0 | 3 | 14 |
| severe | M04 | 84 | 7.9 | 8.0 | 3 | 14 |

Fig. 6 shows that substitutions were the largest error category for every system at severe dysarthria, accounting for between 58% and 74% of per-utterance errors. The systems differed mainly in whether they produced more or less text than was spoken: the OpenAI multimodal models tended to under-generate, with median hypothesis-to-reference length ratios of about 0.81 to 0.83, whereas Gemini 2.5 Pro over-generated, with a median ratio of about 1.13. The conventional ASR systems and Gemini 2.5 Flash were close to parity by median, with Gemini 2.5 Flash showing extreme over-generation only on a minority of repeat-loop utterances. In the M02 under-generation example in Table 6, the omitted words are predominantly content-bearing, such as "misguided" and "way", so the hypothesis remains fluent and grammatical while shedding the words that carry most of the propositional content, rather than dropping only function words or inflectional morphemes.

Fig. 7 further shows that this tendency was not uniform across the four severe speakers. Gemini over-generation was carried mainly by the most affected speaker (M04), whereas the least affected severe speaker (M02) remained at length parity across all systems. We therefore treat this as a tendency in the most degraded speech rather than an architecture-level rule. Given the default decoding configurations used throughout, it may also partly reflect provider-specific generation verbosity.

Whether reduced vocal intensity raises deletion rates in the most affected speakers cannot be tested directly here, because TORGO does not provide per-speaker loudness or sound-pressure-level measurements. The length ratios in Fig. 7 serve only as a coarse proxy for net under-generation. They show a deletion-leaning profile for F01 and, less consistently, M01 under the OpenAI multimodal models, length parity for M02 across all systems, and a mixed pattern for M04, which under-generated under the OpenAI multimodal models, with ratios of about 0.63 to 0.67, yet over-generated under the Deepgram and Gemini systems, with ratios of roughly 1.09 to 1.33, where its high WER was driven by insertions and repetition rather than deletions. The balance between deletions and insertions therefore appears to be jointly determined by speaker and system rather than by speaker severity alone, and a dedicated acoustic analysis incorporating intensity measurements would be required to attribute any deletion tendency specifically to reduced vocal effort. We flag this per-speaker acoustic and intensity analysis as a direction for future work.

Two qualitative failure modes are relevant here. The OpenAI multimodal models occasionally produced fluent output in a non-target language on severe speech, while Gemini 2.5 Flash occasionally produced degenerate repetition, in one case emitting approximately 4,019 words for an eight-word reference. Such extreme cases are why error rates are computed per utterance and capped rather than pooled across words, which would otherwise allow a single output to dominate the aggregate.

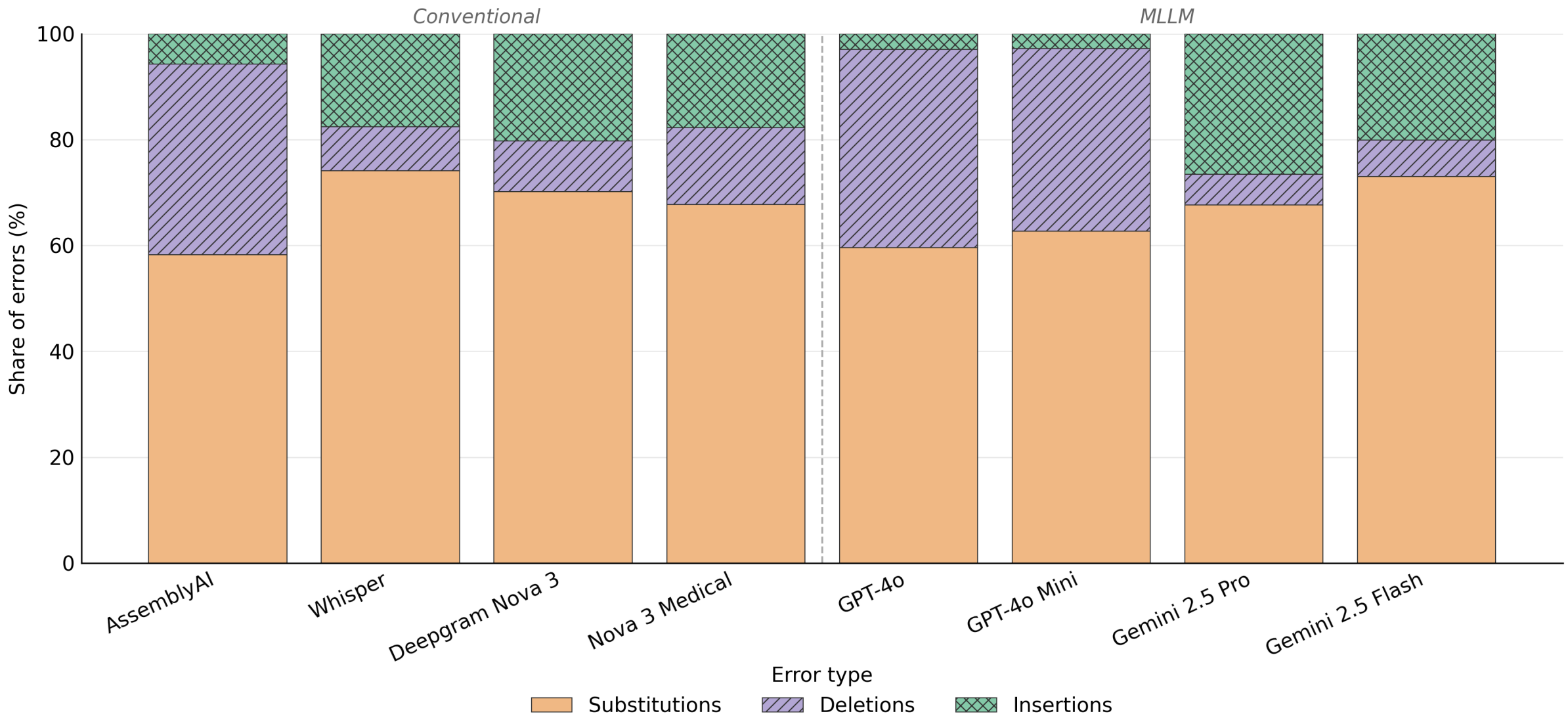

**Fig. 6** Error composition at severe dysarthria, per system, shown as the mean per-utterance share of substitutions, deletions, and insertions under the no-prompt baseline. Shares sum to 100%; absolute error magnitudes are given in Table 3.

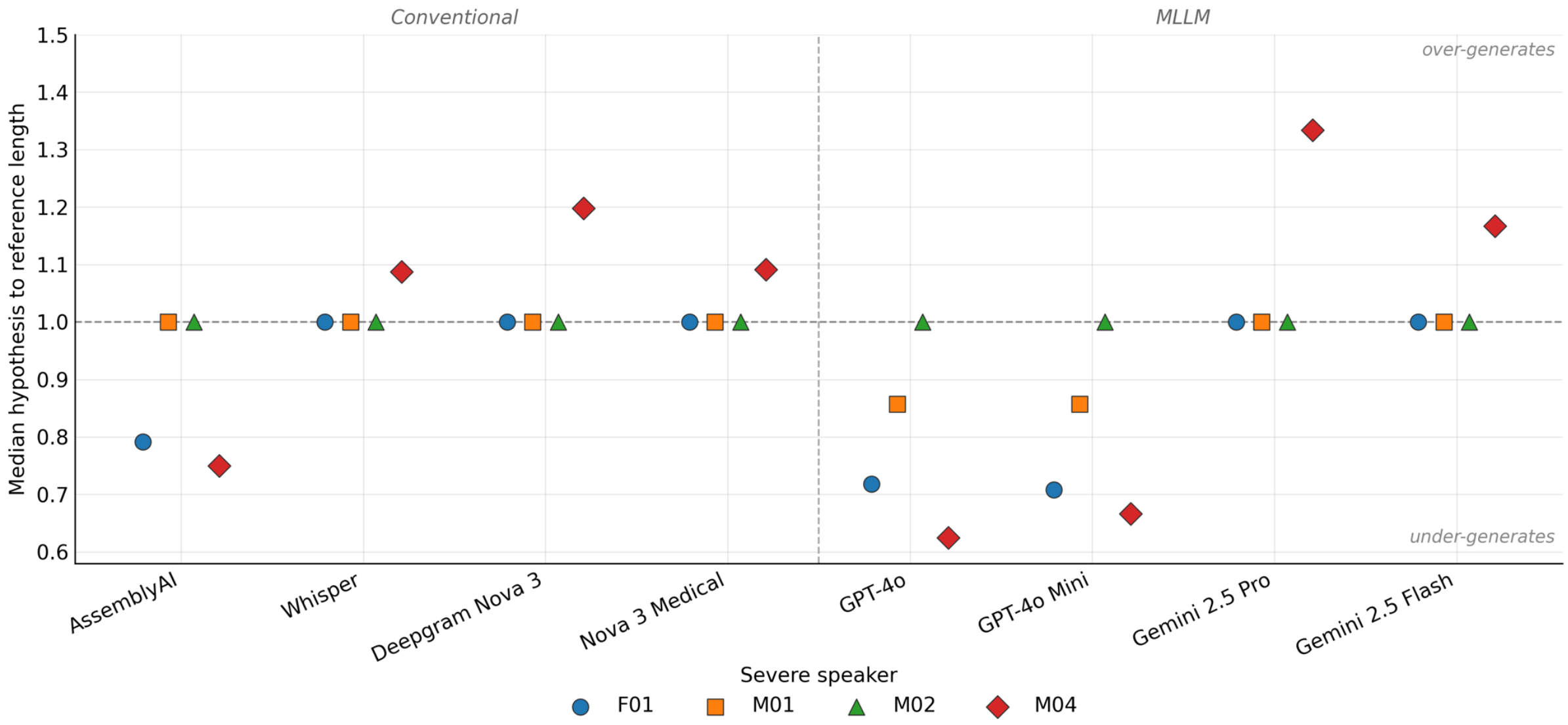

**Fig. 7** Median hypothesis to reference output-length ratio at severe dysarthria, per system, with the four severe speakers shown separately (no-prompt baseline). Values below 1 indicate under-generation and above 1 over-generation; the Gemini over-generation is carried mainly by speaker M04.

WER and the semantic metrics were strongly negatively correlated at the utterance level. Pooled across systems, the Spearman correlation between WER and BERTScore was rho = -0.95, indicating that the semantic metrics are largely redundant with WER in aggregate. SeMaScore behaved similarly and was closely correlated with BERTScore, so the two semantic measures conveyed similar information. Fig. 8 illustrates the lexical-semantic relationship using BERTScore, the semantic metric foregrounded throughout this study.

At the per-severity level, the association was strongest at moderate and severe dysarthria (BERTScore rho = -0.93 and -0.90 respectively) and weakest at mild dysarthria (rho = -0.62). The weaker mild-tier correlation reflects range restriction rather than metric divergence, since most mild utterances were recognised at or near WER = 0, leaving little spread for a rank correlation to act on. Across all severities, a high-error tail was nonetheless present, with 23 of the 1,857 utterances with WER ≥ 50% (1.2%) retaining a BERTScore ≥ 70%, indicating cases in which lexical accuracy is poor yet communicative intent is partly preserved. Because each utterance contributes one observation per system, these pooled observations are not independent; therefore, the correlation coefficients are interpreted descriptively.

Representative cases in which WER overstates communicative loss, alongside contrasting cases of genuine meaning loss, are shown in Table 6. These examples support the use of semantic metrics as complementary indicators, particularly because semantic metrics have been shown to align more closely with human judgements of intelligibility on disordered speech than WER alone [31, 41].

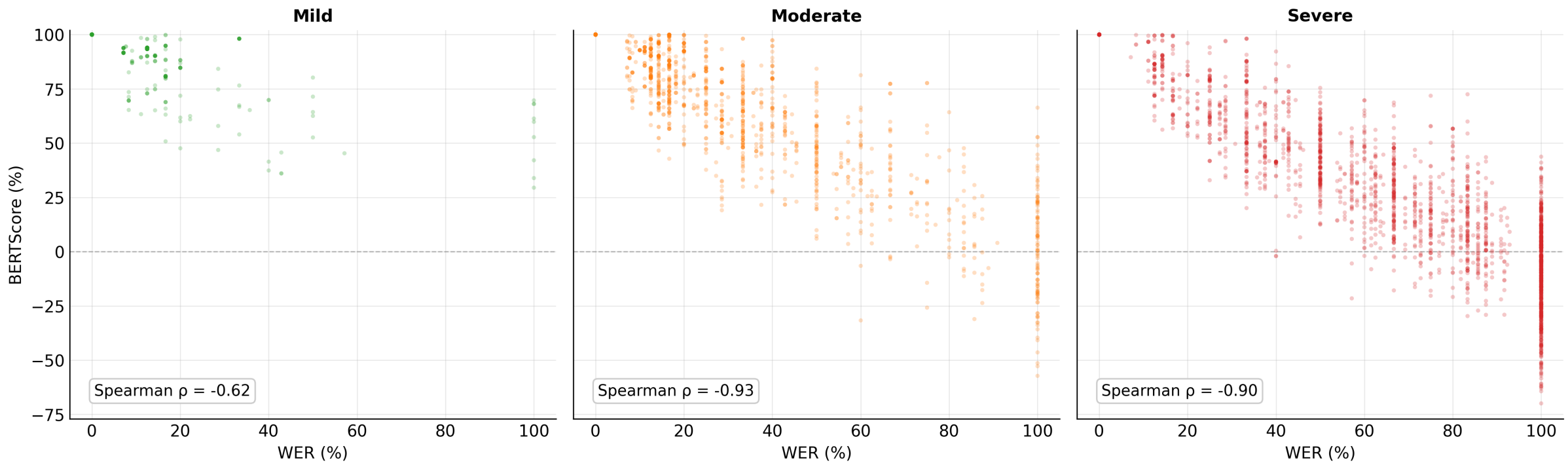

**Fig. 8** Utterance-level WER versus BERTScore by severity (no-prompt baseline), with per-panel Spearman rho. Pooled rho = -0.95.

**Table 6** Representative no-prompt transcription examples showing how WER reflects formatting-only matches, lexical variation, generation errors, and meaning loss after text normalisation.

| Severity | Speaker | System | Category | Reference | Hypothesis | WER | BERTScore |
|---|---|---|---|---|---|---|---|
| Moderate | F03 | Whisper | Formatting only | Well, he is nearly ninety-three years old. | Well, he's nearly 93 years old. | 0.0 | 100.0 |
| Moderate | M05 | Whisper | Formatting only | but he always answers, "Banana oil!" | But he always answers, banana oil. | 0.0 | 100.0 |
| Moderate | F03 | Whisper | Lexical variation, partial meaning preserved | She wore warm, fleecy, woolen overalls. | Okay. She wore warm fleece overalls. | 50.0 | 57.6 |
| Severe | M02 | GPT-4o | Under-generation (deletions) | The misguided souls have lost their way. | The soldiers have lost. | 57.1 | 35.1 |
| Severe | M04 | Gemini 2.5 Pro | Over-generation (insertions) | They carried me off on the stretcher. | They they came me on, you know, on and on and on and on, I'm a b- b- veteran. | 100.0 | -5.4 |
| Severe | F01 | GPT-4o | Meaning loss | Usually minus several buttons. | Lygau mae sawl bacyns. | 100.0 | -26.0 |

## 4.3 Prompting effects on MLLM performance

Having characterised the baseline error and semantic behaviour, we next examine whether a verbatim-transcription prompt improves multimodal recognition, using the four conditions defined in Section 3.3. Fig. 9 and Table 7 show that the effect of prompting was strongly asymmetric across model families. Prompting improved the OpenAI multimodal models at severe dysarthria but did not help the Gemini models at any severity.

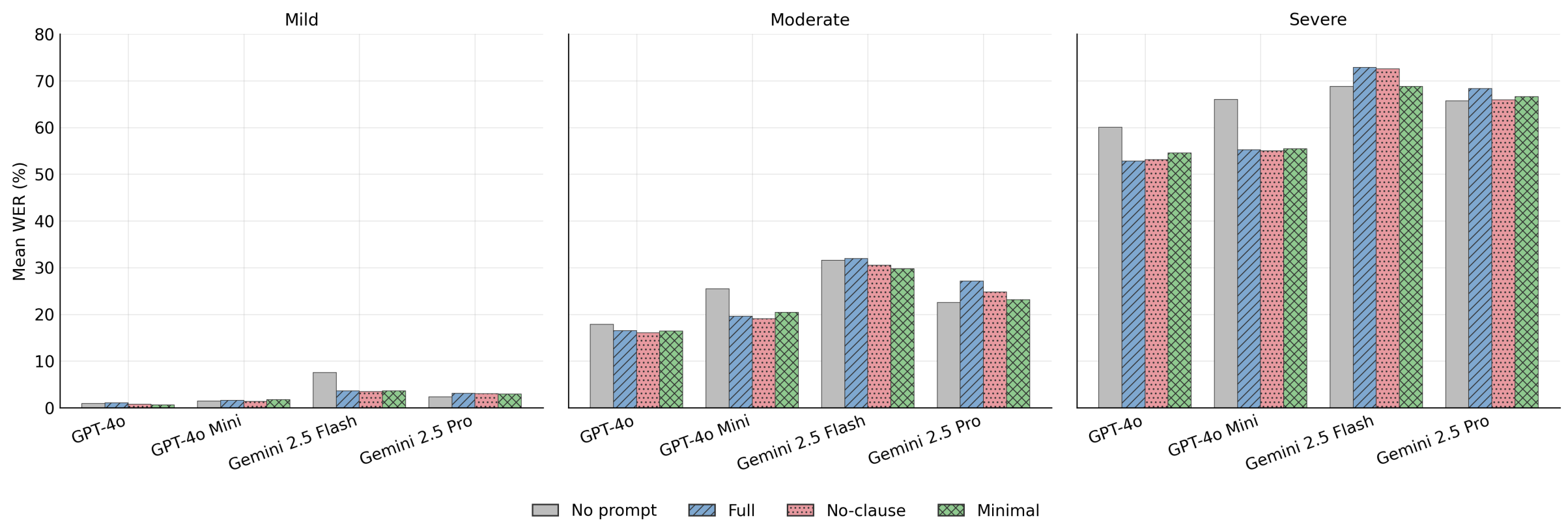

**Fig. 9** Mean WER (%) by prompt condition, model and severity for the four multimodal systems (lower is better). The four conditions are the no-prompt baseline and the full, no-clause and minimal verbatim prompts.

**Table 7** Mean WER (%) for the four multimodal systems across the four prompt conditions, by severity (lower is better). For each model and severity the lowest value is shown in bold.

| Model | Severity | No prompt | Full | No-clause | Minimal |
|---|---|---|---|---|---|
| GPT-4o | mild | 0.87 | 1.03 | 0.77 | **0.62** |
| GPT-4o | moderate | 17.85 | 16.52 | **16.05** | 16.43 |
| GPT-4o | severe | 60.08 | **52.86** | 53.14 | 54.54 |
| GPT-4o Mini | mild | 1.40 | 1.55 | **1.32** | 1.74 |
| GPT-4o Mini | moderate | 25.45 | 19.57 | **19.06** | 20.39 |
| GPT-4o Mini | severe | 66.00 | 55.27 | **55.01** | 55.49 |
| Gemini 2.5 Flash | mild | 7.56 | 3.58 | **3.50** | 3.62 |
| Gemini 2.5 Flash | moderate | 31.57 | 31.96 | 30.50 | **29.77** |
| Gemini 2.5 Flash | severe | 68.83 | 72.87 | 72.56 | **68.80** |
| Gemini 2.5 Pro | mild | **2.34** | 3.12 | 3.03 | 2.96 |
| Gemini 2.5 Pro | moderate | **22.50** | 27.09 | 24.81 | 23.15 |
| Gemini 2.5 Pro | severe | **65.74** | 68.34 | 65.92 | 66.64 |

For GPT-4o, Table 7 shows that the severity-level mean WER at the severe tier fell from 60.1% without a prompt to between 52.9% and 54.5% under the three prompted conditions; for GPT-4o Mini, it fell from 66.0% to as low as 55.0%. These are means across the four severe speakers; Table 8 shows that individual speakers varied more widely. The Gemini models showed no comparable benefit at severe dysarthria and were, if anything, slightly worse under prompting.

The four-condition design also allows the contribution of the anti-inference clause to be examined directly, since the full and no-clause prompts differ only by that single sentence. For GPT-4o Mini at severe dysarthria, mean WER was 55.3% under the full prompt, 55.0% under the no-clause prompt and 55.5% under the minimal prompt. The three prompted conditions therefore differed by less than one percentage point and were all close to one another, well below the roughly eleven-point gain each delivered over the 66.0% no-prompt baseline. The same pattern held for GPT-4o, whose three prompted conditions at severe dysarthria fell within about two points of each other, from 52.9% to 54.5%. The benefit of prompting for the OpenAI multimodal models is therefore associated with the presence of any verbatim-transcription instruction rather than with the anti-inference clause specifically, and we do not claim an isolated effect for that clause.

This pattern is consistent with the failure-mode mechanism identified above. Without a prompt, the OpenAI multimodal models produced fluent non-target-language output on a substantial minority of severe utterances, 38 of 284 for GPT-4o and 54 of 284 for GPT-4o Mini. By contrast, the Gemini models produced such outputs at most twice in any condition. Fig. 10 shows that, under each prompted condition, OpenAI language drift fell sharply, to at most eleven utterances.

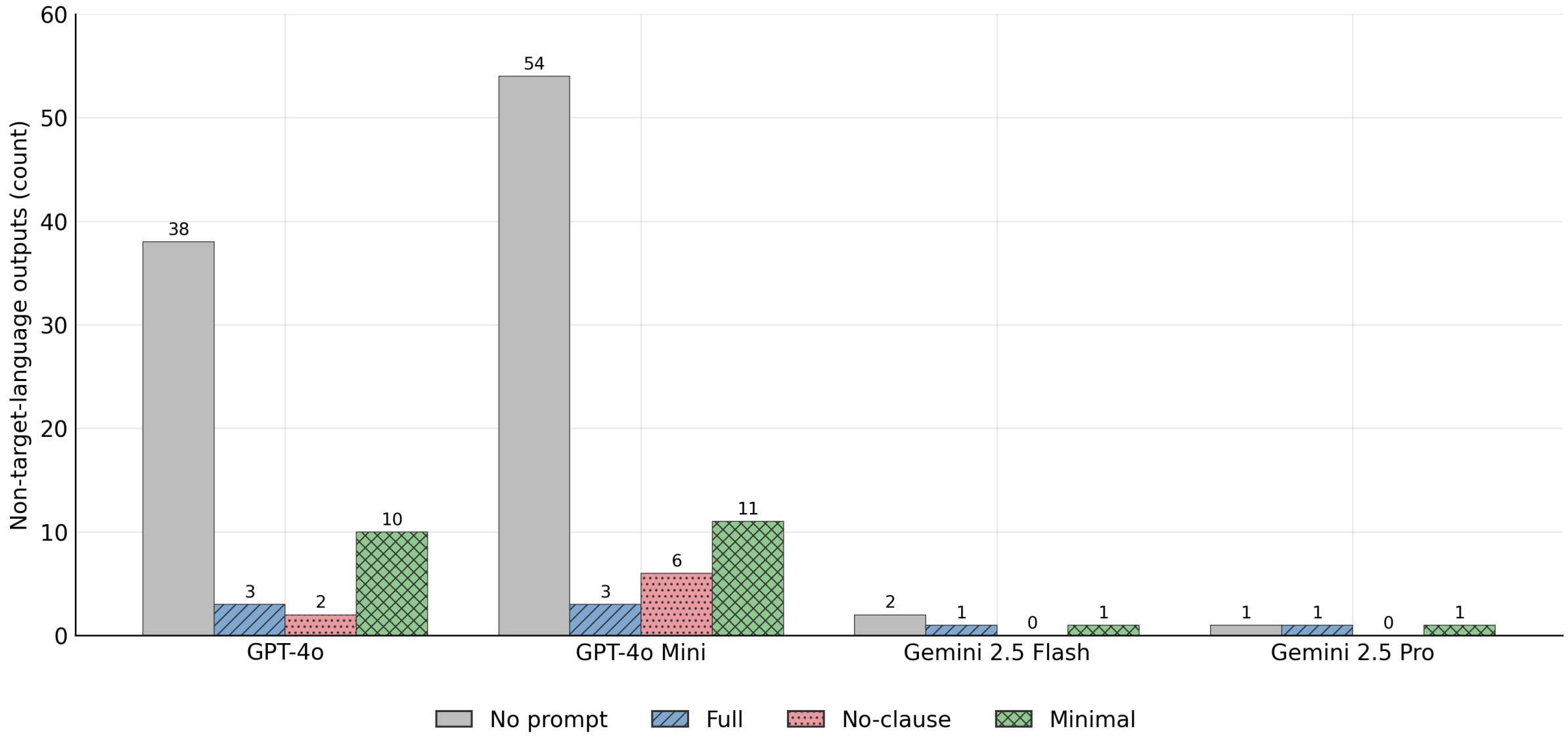

**Fig. 10** Non-target-language outputs per multimodal model across the four prompt conditions at severe dysarthria (out of 284 severe utterances). Counts are based on automatic detection and are approximate.

The verbatim English-transcription instruction therefore appears to counter a failure mode that is more prominent in the OpenAI multimodal models than in the Gemini models, providing a plausible explanation for why prompting helps the former but not the latter.

Two further observations qualify the practical recommendation. First, for the OpenAI multimodal models the three prompted variants performed similarly, with the full, no-clause and minimal prompts differing by about one percentage point at severe dysarthria, so a short verbatim-transcription instruction was sufficient and the longer prompt offered no consistent advantage. This indicates that prompt simplification can be useful, but that the benefit of prompting at all remains model-specific rather than universally transferable, since the Gemini models did not respond to any of the prompted variants.

Second, prompting can harm less impaired speech. For the Gemini models, prompting produced no consistent benefit at mild or moderate severity and sometimes increased WER; for example, Gemini 2.5 Pro, which was the strongest Gemini configuration without a prompt at every severity, rose from 22.5% to 27.1% mean WER at moderate dysarthria under the full prompt. We therefore recommend prompting selectively, at severe dysarthria and for models that exhibit language drift, rather than as a default. It should also be emphasised that prompting is a partial mitigation rather than a solution: residual severe-tier WER remained above 50% for all multimodal systems and prompt conditions, and the analysis rests on only four severe speakers.

Table 8 and Fig. 11 show that the OpenAI improvement at severe dysarthria was not driven by a single speaker: mean WER fell under prompting for GPT-4o and GPT-4o Mini across all four severe speakers (F01, M01, M02, and M04). The Gemini models showed no comparable consistency.

Gemini 2.5 Pro improved on F01 under all three prompts, but this gain was offset by losses on other speakers. Because F01 contributes only 20 utterances, it carries little weight in the utterance-weighted aggregate reported in Table 7, where the Gemini models remain flat or slightly worse under prompting.

**Table 8** Per-speaker mean WER (%) for the four multimodal systems across the four prompt conditions, at severe dysarthria (lower is better). The improvement under prompting holds across all four severe speakers for the OpenAI multimodal models, whilst the Gemini models show no consistent improvement.

| Speaker | Model | No prompt | Full | No-clause | Minimal |
|---|---|---|---|---|---|
| F01 | GPT-4o | 61.83 | **50.76** | 62.72 | 53.17 |
| F01 | GPT-4o Mini | 72.54 | 60.63 | 60.49 | **57.72** |
| F01 | Gemini 2.5 Flash | **60.16** | 64.81 | 63.20 | 63.13 |
| F01 | Gemini 2.5 Pro | 59.51 | 62.65 | 62.28 | **55.84** |
| M01 | GPT-4o | 57.96 | 48.10 | **46.19** | 48.12 |
| M01 | GPT-4o Mini | 62.53 | 49.33 | **48.36** | 49.31 |
| M01 | Gemini 2.5 Flash | 65.21 | 68.72 | 68.81 | **64.46** |
| M01 | Gemini 2.5 Pro | 63.75 | 65.86 | **61.50** | 64.17 |
| M02 | GPT-4o | 47.66 | **45.32** | 45.43 | 47.00 |
| M02 | GPT-4o Mini | 53.02 | 45.82 | **45.19** | 48.63 |
| M02 | Gemini 2.5 Flash | 57.74 | 62.24 | 61.25 | **56.58** |
| M02 | Gemini 2.5 Pro | **50.18** | 53.85 | 51.94 | 53.46 |
| M04 | GPT-4o | 75.37 | **66.55** | 66.57 | 69.85 |
| M04 | GPT-4o Mini | 82.18 | 70.53 | 71.39 | **68.95** |
| M04 | Gemini 2.5 Flash | **86.75** | 90.71 | 91.02 | 87.98 |
| M04 | Gemini 2.5 Pro | 86.20 | 88.03 | 86.62 | **86.09** |

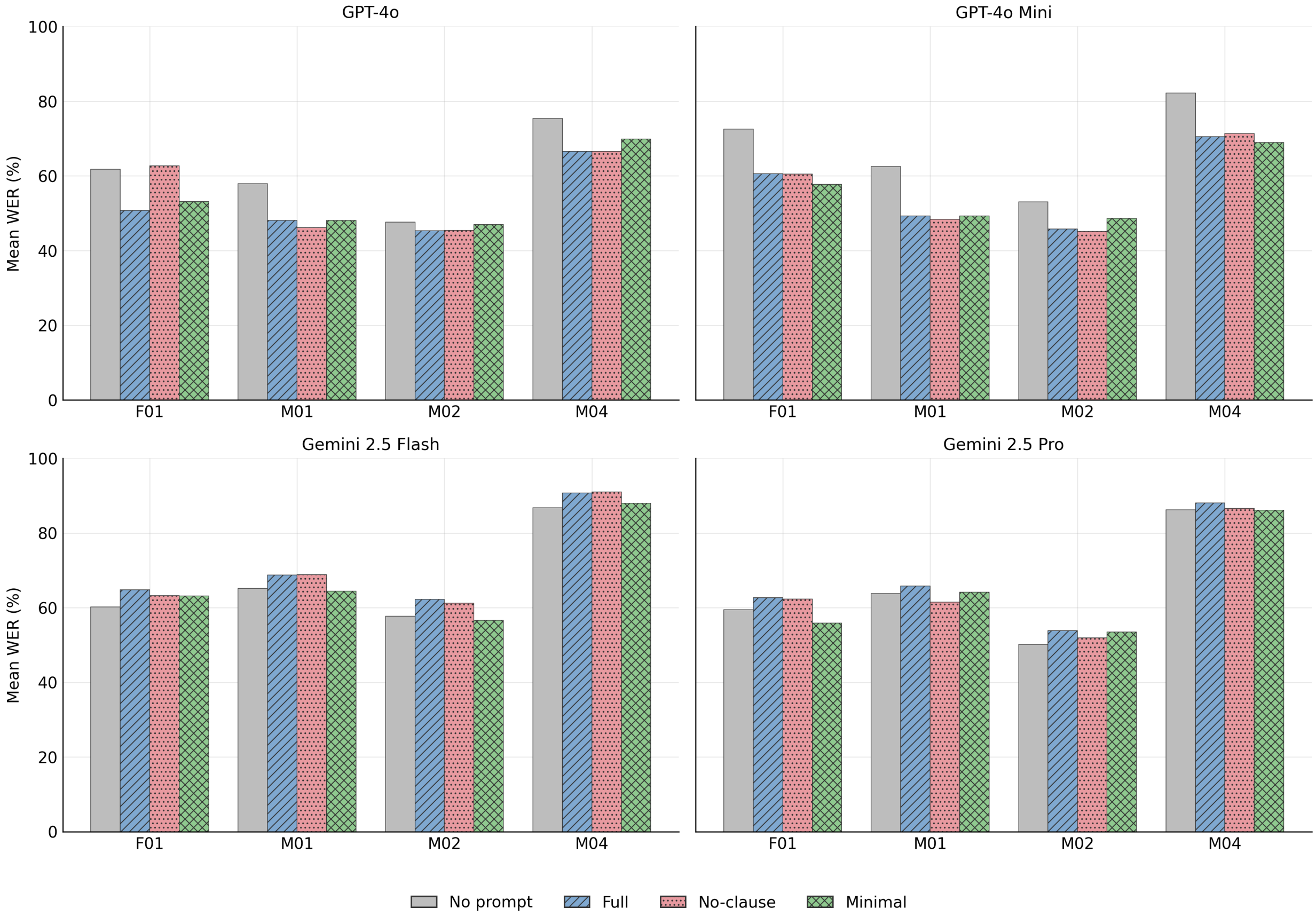

**Fig. 11** Per-speaker mean WER (%) by prompt condition for the four multimodal systems at severe dysarthria, shown for the four severe speakers (F01, M01, M02, M04; lower is better). Conditions as in Fig. 9.

Fig. 12 shows that the same asymmetry extends to the primary semantic metric. BERTScore improved for the OpenAI multimodal models at severe dysarthria under prompting, whereas it changed little or decreased for the Gemini models, mirroring the WER pattern.

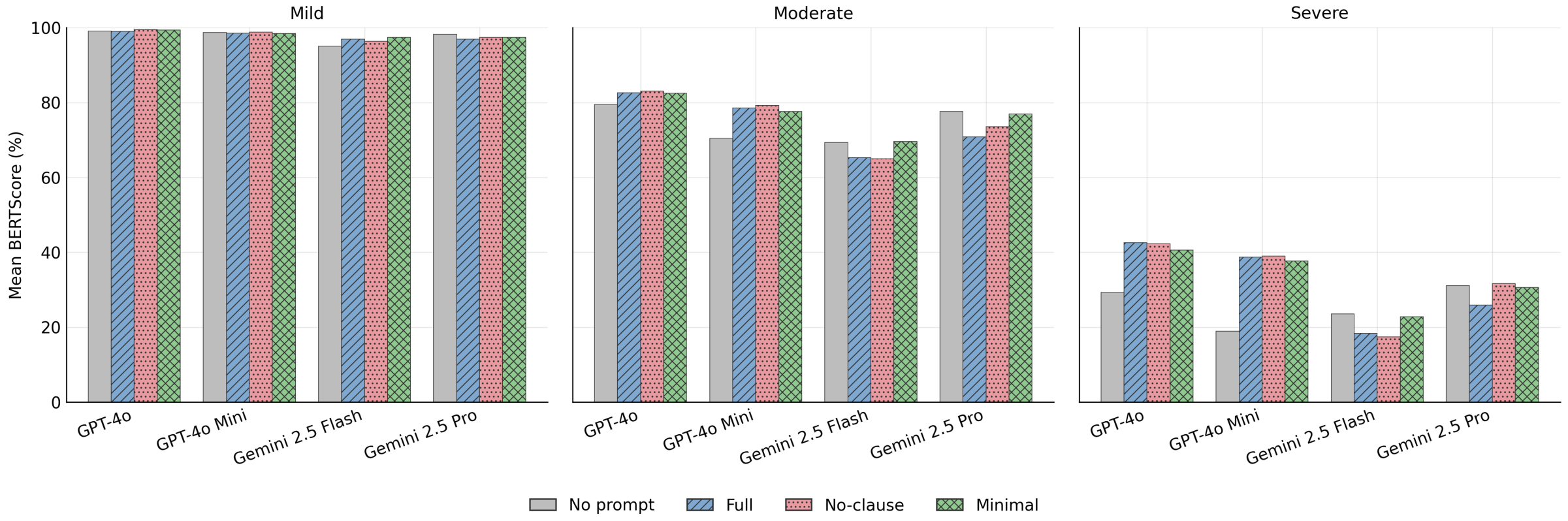

**Fig. 12** Mean BERTScore (%) by prompt condition, model and severity for the four multimodal systems (higher is better). Conditions as in Fig. 9.

## 4.4 Cost and latency trade-offs

Operational characteristics varied substantially across systems. These differences are summarised visually in Fig. 13 and numerically in Table 9. The cost and latency values describe general service characteristics rather than properties specific to dysarthric input. Because the prompt conditions differ by only a small number of input tokens, prompt-related cost differences are expected to be small; latency differences are reported descriptively because they may also reflect API load, network conditions, and provider-side processing behaviour. We therefore report these operational measures at the system level to inform deployment feasibility for this population.

For context, human conversational studies suggest typical response latencies around 300 to 500 ms; delays beyond approximately 500 ms feel noticeably unnatural. In designing real-time AAC and voice interfaces, this range is often used as a target for natural interaction, though acceptable thresholds vary by application context and individual user expectations [42]. Asynchronous applications, such as message composition, can tolerate considerably longer processing times.

Conventional ASR services occupied the efficient region of the cost-latency trade-off space, combining low cost with predictable processing times. Deepgram Nova-3 achieved US$0.124 per minute at 688 seconds total corpus processing time, whilst Whisper demonstrated similar efficiency (US$0.173 per minute, 518 seconds). AssemblyAI, despite strong accuracy performance, incurred higher cost and latency (US$0.226 per minute, 1,741 seconds).

MLLM-based systems exhibited more variable efficiency profiles. Non-prompted Gemini variants offered the lowest absolute costs (Flash: US$0.055 per minute; Pro: US$0.069 per minute), but with substantially elevated latency (943 and 2,765 seconds, respectively). Prompting increased Gemini latencies further, with Pro rising to 5,470 seconds, without corresponding accuracy benefits. GPT-4o, by contrast, demonstrated more favourable prompting efficiency, with latency decreasing from 780 to 647 seconds under prompting, whilst cost increased moderately from US$0.260 to US$0.303 per minute.

Given that typical AAC utterances are 2 to 5 seconds in duration, per-utterance costs would be approximately 3 to 8% of the per-minute figures reported, placing even premium services within practical budgets for individual use. The latency estimates reported in Table 9 should be interpreted as proxy values derived from aggregate processing times; real-world latency will vary with client location, network conditions, and API load. The relative rank ordering of systems is likely more stable than the absolute millisecond values reported.

**Table 9** Speaker-level estimated per-utterance latency (milliseconds). Median, interquartile range (IQR), and range computed across speaker-level estimates (n = 8 speakers).

| System | Cond. | Median | IQR (Q25-Q75) | Min-Max |
|---|---|---|---|---|
| GPT-4o Mini | NP | 1,567 | 1,462-1,733 | 1,431-2,138 |
| Whisper | NP | 2,150 | 2,071-2,428 | 1,705-2,527 |
| Gemini 2.5 Flash | NP | 2,452 | 1,630-3,315 | 1,528-15,949 |
| Deepgram Nova 3 Medical | NP | 2,480 | 2,210-2,780 | 1,980-3,520 |
| Deepgram Nova 3 | NP | 2,691 | 2,384-3,009 | 2,181-3,814 |
| GPT-4o | NP | 3,197 | 2,913-3,383 | 2,816-3,990 |
| AssemblyAI | NP | 6,978 | 6,775-7,368 | 6,279-7,820 |
| Gemini 2.5 Pro | NP | 13,103 | 5,483-18,317 | 3,391-36,148 |
| GPT-4o Mini | P | 1,210 | 1,067-1,429 | 1,004-1,886 |
| GPT-4o | P | 2,319 | 1,962-2,993 | 1,797-3,402 |
| Gemini 2.5 Flash | P | 6,204 | 2,621-10,081 | 2,138-30,352 |
| Gemini 2.5 Pro | P | 17,148 | 7,839-26,544 | 4,913-39,027 |

*Note: NP = non-prompted. P = prompted condition used for the operational comparison, corresponding to the full verbatim-transcription prompt. Latency statistics were derived from per-speaker aggregate processing times divided by utterance counts. Median and IQR reflect the distribution of these speaker-level estimates (n = 8 speakers), and Min-Max indicates the range across speaker-level means rather than individual utterance extremes. The 300-500 ms conversational latency range discussed in Section 4.4 is not achieved by any evaluated system; all systems are therefore better suited to asynchronous or confirmation-based use cases where processing delays are tolerable.*

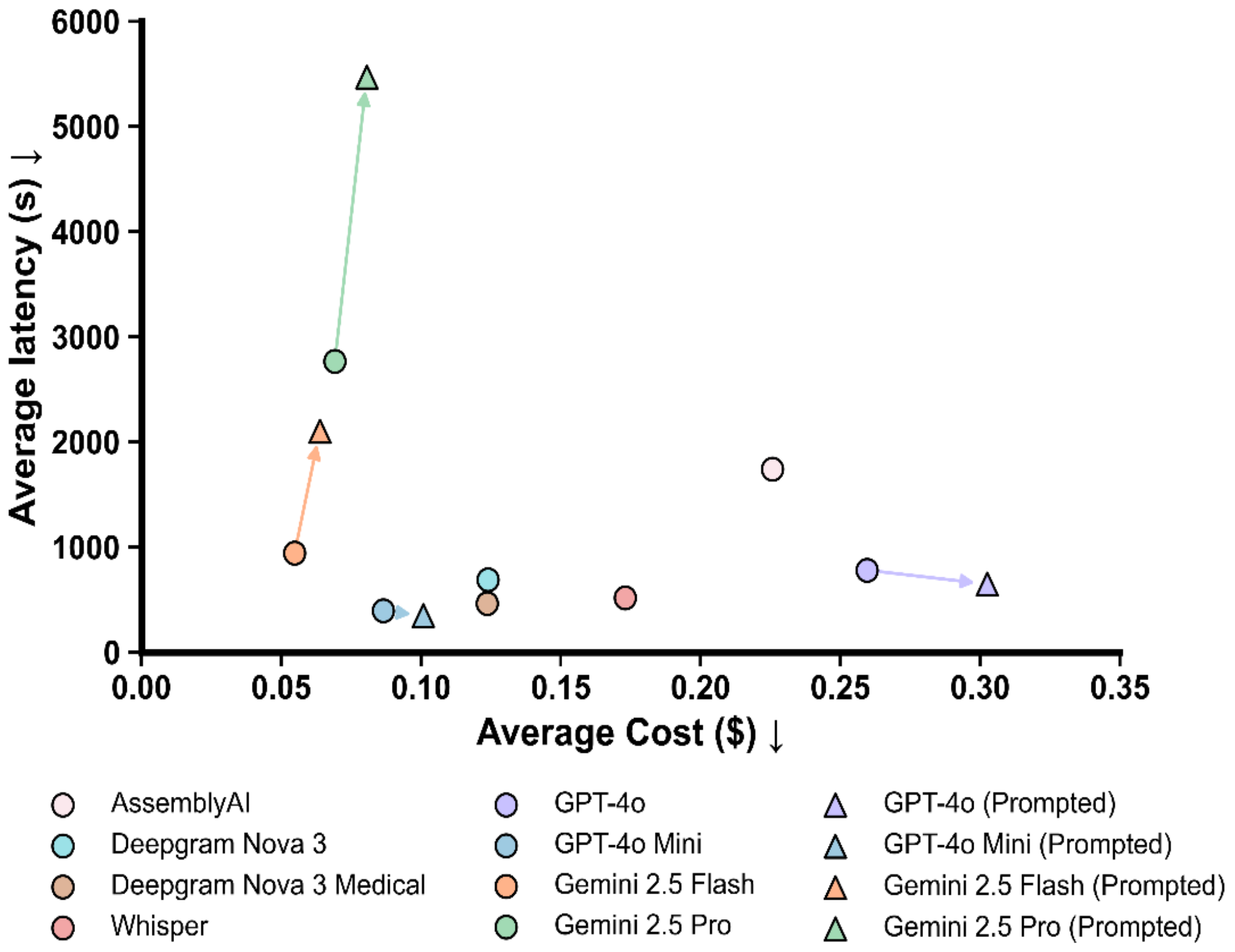

**Fig. 13** Cost and latency characteristics of evaluated systems. Points show average cost (US dollars per minute) versus total corpus processing time (seconds) for non-prompted and prompted conditions.

# 5 Discussion

## 5.1 Implications for universal access and inclusive design

The findings illuminate both the current state and future direction of inclusive voice interface design. The severity-dependent performance gradient shows that commercial speech recognition systems, designed primarily around typical speech patterns, continue to disadvantage users whose speech characteristics fall outside assumed acoustic norms. This exclusion is not merely a technical limitation, but also reflects design choices that privilege acoustic homogeneity over speech diversity. From the perspective of design for all [6, 14], current voice technologies therefore demonstrate limited success in proactively accommodating the full range of human speech variation.

However, the results also identify pathways toward more inclusive design. For users with mild dysarthria, multiple systems achieved recognition accuracy approaching typical-speech benchmarks, suggesting that voice interfaces can already serve this population with minimal accommodation. For moderate dysarthria, semantic preservation metrics indicate that user intent may remain recoverable even when lexical accuracy degrades, supporting interface designs that incorporate confirmation mechanisms, semantic disambiguation, or intelligent error correction. Table 6 provides representative examples illustrating this distinction between lexical error and preserved communicative meaning. For severe dysarthria, the substantial performance gap, exceeding 51% WER across all systems, underscores the continuing need for alternative and multimodal interaction strategies, with voice functioning as one input among several rather than as the sole modality.

The architecture-specific prompting effects observed in this study have direct implications for assistive technology practice. At severe dysarthria, prompting should be interpreted as a model-specific mitigation rather than as a general accessibility intervention. For the OpenAI multimodal models, GPT-4o and GPT-4o Mini, verbatim-transcription prompting reduced severe-tier WER by suppressing non-target-language drift, with both improving under any verbatim-transcription prompt and the three prompt variants performing similarly. This suggests that prompting may be useful when repeated severe-speech failures, language drift, or over-correction are observed, but it should not be applied automatically across all models or severity levels. For Gemini-based systems, prompted variants did not improve accuracy and sometimes degraded performance. This divergent response may reflect architectural differences in how these systems balance acoustic evidence against linguistic inference. The broader implication is that accessibility interventions cannot be assumed to generalise across technological platforms and must be validated empirically.

## 5.2 Practical and design implications for assistive deployment

Drawing on our empirical findings, we offer consolidated guidance for practitioners selecting and configuring speech recognition for assistive applications. This section integrates deployment considerations (application type, cost, latency) with severity-stratified design recommendations aligned with design for all principles.

*For users with mild dysarthria*, current commercial systems can support voice-primary interfaces with minimal accommodation. Recognition accuracy (around 1 to 2% WER) approaches typical-speech performance. GPT-4o or Whisper in non-prompted mode is therefore recommended for this severity level. Interface design can follow typical voice-assistant patterns, supported by standard confirmation and correction options. The high benchmarked semantic-preservation scores indicate that communicative intent is usually retained at this severity level. In deployed systems, standard ASR confidence scores, user confirmation, and correction options should be used to manage residual errors.

*For users with moderate dysarthria*, voice interfaces should incorporate explicit confirmation and correction mechanisms. Based on the observed WER values, Whisper and non-prompted GPT-4o are the preferred configurations for lexical accuracy, whereas GPT-4o Mini with a verbatim-transcription prompt may be considered when an MLLM-based workflow is required. Prompting should not be applied automatically at this severity level because the benefit was model- and severity-dependent rather than universal. Interface design should present transcription results with confidence indicators, alternative hypotheses where available, and one-tap correction options. The benchmarked BERTScore values should be interpreted as retrospective evidence that semantic intent can remain partly recoverable despite elevated lexical error rates; however, because BERTScore requires reference transcriptions, it should not be treated as a direct runtime threshold. In deployed systems, this finding instead motivates runtime proxies such as ASR confidence scores, N-best agreement, self-consistency checks, and explicit user confirmation. Alternative input modalities, including on-screen keyboard input and predictive text, should also be readily available as fallback options.

*For users with severe dysarthria*, voice should function as one component within a multimodal interaction strategy rather than as a standalone modality. Among the MLLM configurations, prompted OpenAI multimodal models provided the strongest improvement at this severity level. GPT-4o under the full prompt achieved the lowest WER among the prompted MLLM conditions, while GPT-4o Mini also improved substantially across the four severe speakers. Selection between these systems should therefore depend on whether lexical accuracy, semantic preservation, latency, or integration constraints are prioritised.

Although the present evaluation tested transcription accuracy rather than interface design, the following directions, informed by these results together with the design-for-all and AAC literature [2, 13, 14], are likely to be productive rather than being conclusions established by the present study: (i) multimodal input combining voice with touch, switch scanning, or eye gaze; (ii) partner-assisted confirmation workflows where a communication partner can verify intent; (iii) context-aware intent inference using conversation history and situational cues to disambiguate uncertain transcriptions; and (iv) progressive correction options, ranging from simple accept/reject confirmation to detailed word-level editing. Given the high residual WER at this severity level, even under prompting, all outputs should be treated as provisional and should be confirmed before being used in safety-critical, clinical, or high-consequence communication.

*Application-type considerations:* For real-time, interactive applications, including AAC conversational interfaces, voice-controlled environmental systems, and real-time captioning, none of the evaluated systems meet the conversational latency targets discussed in Section 4.4. Among the available options, GPT-4o Mini and Whisper offer the most practical latency profiles, although both remain better suited to short asynchronous interactions than fully natural conversational turn-taking. For asynchronous, intent-critical applications such as message composition, email preparation, and document dictation, system choice should be severity-dependent. For mild and moderate dysarthria, non-prompted GPT-4o or Whisper should be preferred. For severe dysarthria, prompted OpenAI variants should be considered, with GPT-4o under the full prompt offering the lowest WER and GPT-4o Mini providing a lower-cost alternative. For resource-constrained deployments, Gemini Flash offers the lowest cost, but its accuracy limitations and negative response to prompting restrict its suitability to lower-risk applications.

*Engine selection decision flow:* When user severity is unknown at design time, systems should begin with a non-prompted high-performing configuration such as GPT-4o or Whisper and adapt based on observed recognition behaviour. Prompting should be enabled selectively when repeated severe-speech failures, non-target-language drift, or over-correction are observed, rather than applied as a default across all users or systems. Prompted Gemini variants should be avoided under the tested conditions because they did not show consistent accuracy benefits and sometimes degraded performance. Interfaces should adjust confirmation requirements, fallback options, and correction support based on observed recognition reliability over time. These findings are most directly applicable to open-vocabulary dictation and phrase-level AAC scenarios; generalisability to short-command smart-home control or constrained-vocabulary applications is limited given the exclusion of single-word utterances from the present evaluation.

## 5.3 Limitations and future work

Several limitations constrain the generalisability of these findings and highlight directions for future research.

*Corpus limitations:* The TORGO corpus, whilst established as a dysarthric speech benchmark, comprises a relatively small participant sample (n = 8 dysarthric speakers) and a single language variety, namely North American English. Its severity classifications are corpus-provided rather than derived from standardised clinical assessments, which limits comparability with clinical severity scales used in practice. The mixed aetiology sample, including speakers with cerebral palsy and ALS, may also obscure subtype-specific recognition patterns. In addition, utterance counts were unequal across speakers and TORGO does not present an identical prompt set to each speaker, so between-speaker comparisons may partly reflect differences in spoken content rather than severity alone; the severity-level comparisons, which aggregate over speakers, are correspondingly less sensitive to individual speakers than the per-speaker contrasts.

Cross-corpus validation using additional dysarthric speech datasets would therefore strengthen the generalisability of the findings. However, direct comparison is constrained by differences in corpus design. UA-Speech consists exclusively of isolated single words, with no sentence-level material [35], whereas the present study focuses on connected utterances of three or more words. The multi-word material in Nemours comprises fixed-frame, semantically anomalous sentences, together with two standard read passages [43]; the former are unsuitable for semantic-preservation analysis because their content is deliberately meaningless and structurally invariant, and the latter are too few to support corpus-level semantic evaluation. The Speech Accessibility Project (SAP) is a more recent and substantially larger English resource that does contain meaningful connected speech across multiple aetiologies [44]; we did not incorporate it here because access is gated behind an institutional data-use agreement and application-approval process that precluded inclusion within this study's timeframe. For these reasons, TORGO was retained as the primary evaluation corpus, and connected-speech cross-corpus validation, with SAP as the natural candidate, is treated as an important direction for future work.

The substantial within-severity variability observed in the per-speaker analysis further indicates that future studies should include larger, more diverse samples spanning multiple languages, aetiologies, and clinically validated severity assessment frameworks.

*Absence of user-centred validation:* This study employed quantitative performance metrics without direct input from dysarthric users. Whilst WER and semantic similarity provide useful technical characterisation, they do not capture the full complexity of functional communication success, user satisfaction, perceived autonomy, or trust in assistive technology. A system that achieves 30% WER might be perceived as highly useful by one user for whom it enables previously impossible communication, yet frustrating for another user with higher accuracy expectations. The practical significance of our findings therefore remains to be validated through user-centred evaluation. We strongly recommend that practitioners consider user feedback alongside technical metrics when selecting systems for deployment.

*Semantic metric validity:* The embedding-based semantic measures employed (SeMaScore, BERTScore) were developed and validated on typical speech transcription tasks, although related ASR-derived intelligibility measures have been clinically validated on dysarthric speech across several conditions and languages [32]. Their behaviour on dysarthric speech, where both reference utterances and recognition hypotheses may contain atypical lexical patterns or disfluencies, has not been extensively validated, and such metrics can over-weight salient keywords [45]. High semantic similarity scores may not correspond to functional communicative success in all AAC contexts, and low scores may underestimate utility for users whose communication partners can interpret context. These metrics should be treated as reasonable proxies pending dedicated validation studies.

*Methodological constraints:* The zero-shot evaluation design, whilst reflecting realistic deployment constraints, does not explore the potential of speaker-adaptive or fine-tuned approaches that may offer superior performance for specific users. Commercial API configurations were fixed at provider defaults, limiting exploration of parameter spaces that might improve accessibility. The prompt-component variants tested do not exhaust the space of possible prompt designs; other speech-relevant approaches, such as few-shot prompting, prompt tuning, domain-specific vocabulary hints, and multi-turn correction, warrant systematic investigation. More reasoning-oriented strategies, such as chain-of-thought prompting or instruction decomposition, were also not examined because verbatim transcription is primarily a perception-bound task rather than a multi-step reasoning task. Nevertheless, the per-speaker consistency of the OpenAI improvements suggests that the tested prompt variants capture a meaningful intervention rather than a single-speaker artefact.

Latency measurements were derived from per-speaker aggregate timing data; whilst this provides reliable central tendency estimates, individual utterance variability may be underestimated. Additionally, this study did not explicitly measure hallucination or unfaithful-generation rates, wherein MLLMs generate plausible but incorrect words. This phenomenon may be particularly relevant for severe dysarthria, where acoustic evidence is weakest and models may rely more heavily on linguistic inference.

*Recording conditions:* All evaluations used TORGO wav_headMic recordings captured with a head-mounted electret microphone in controlled laboratory conditions. Recognition performance may differ with array microphone recordings, ambient noise, or consumer-grade recording equipment typical of real-world AAC deployment. Future evaluations should assess robustness across recording conditions.

*Ethical and privacy considerations:* The use of commercial cloud-based APIs for assistive speech recognition raises important privacy and data governance considerations. Audio data transmitted to third-party servers may be subject to retention, logging, or use for model improvement, with policies varying across providers and potentially changing over time. For clinical or healthcare-adjacent applications, practitioners must verify that selected services comply with applicable regulations (e.g., HIPAA in the United States, GDPR in the European Union) and obtain appropriate user consent for data transmission. Where privacy requirements are stringent, on-device alternatives such as locally deployed Whisper models may be preferable despite potential accuracy trade-offs. The TORGO corpus used in this study is publicly available with appropriate ethical approvals; no new human subjects data were collected.

Future research should address several directions identified by this study. First, systematic investigation of prompting strategies, including few-shot examples, domain-specific instructions, and multi-turn designs, may reveal additional interventions accessible to practitioners. Second, user-centred evaluations linking quantitative performance to perceived usability, trust, and communication success would complement technical assessment and validate the practical significance of observed differences. Third, hybrid approaches combining ASR with downstream semantic processing may bridge the gap between lexical accuracy and communicative adequacy. Fourth, evaluation on more recent and diverse dysarthric speech corpora would strengthen generalisability of findings. Fifth, characterisation of hallucination patterns in MLLM transcription of severely degraded speech would inform risk assessment for safety-critical applications. Sixth, evaluation of short-command recognition using appropriately designed protocols would address the gap created by excluding single-word utterances from the present analysis.

## 6 Conclusion

This study presents a severity-stratified, zero-shot evaluation of commercial speech recognition services for dysarthric speech, addressing an important gap in the universal access evidence base. Eight systems, comprising four conventional ASR and four MLLM-based services, were assessed on the TORGO dysarthric speech corpus across lexical accuracy, semantic fidelity, and operational efficiency. The findings establish empirical baselines that quantify the current accessibility gap: whilst users with mild dysarthria may achieve near-typical voice-interface functionality, those with severe dysarthria continue to face recognition failure rates exceeding 51% WER across all evaluated systems.

The study makes three primary contributions to the universal access literature. First, it provides actionable, severity-stratified performance data to support evidence-based technology selection for assistive applications, including per-speaker analysis demonstrating substantial within-severity variability. Second, it characterises architecture-specific prompting effects through a four-condition ablation, showing that prompting benefits were not general across MLLM systems. The OpenAI multimodal models improved consistently across the four severe TORGO speakers under any verbatim-transcription prompt, mainly through suppression of non-target-language drift, whereas the Gemini models showed no consistent benefit and sometimes degraded. This finding has direct deployment implications: prompting should be treated as a model- and failure-mode-specific intervention rather than as a universal accessibility enhancement.

Third, the study clarifies the role of semantic evaluation in dysarthric ASR benchmarking by showing that SeMaScore and BERTScore are strongly correlated with WER and largely redundant in aggregate, yet still useful for identifying a high-error tail in which lexical accuracy is poor but communicative intent is partly preserved. Together, these findings situate speech recognition accessibility within the broader discourse on inclusive design, demonstrating how commercial voice technologies continue to privilege typical speech characteristics unless atypical speech is explicitly considered during evaluation and deployment.

Accessible voice interaction is not a peripheral concern, but a matter of digital inclusion in an information society increasingly mediated by speech. As voice interfaces proliferate across domestic, occupational, and public domains, the performance gap documented here translates into systematic exclusion of individuals with dysarthria from the benefits of voice-mediated technology. Closing this gap requires continued research, commercial attention to diverse speech populations, and design practices that treat human speech diversity not as an exception, but as a fundamental condition of inclusive voice-interface design.

## Acknowledgements

An earlier version of this article has been made available as a preprint on arXiv.